# Advances and Challenges of SCAN and r²SCAN Density Functionals in Transition-Metal Compounds

Yubo Zhang[1,*], Akilan Ramasamy[2], Kanun Pokharel[2], James W. Furness[2], Jinliang Ning[2], Ruiqi Zhang[2], and Jianwei Sun[2,*]

[1]*Minjiang Collaborative Center for Theoretical Physics, College of Physics and Electronic Information Engineering, Minjiang University, Fuzhou 350108, China*

[2]*Department of Physics and Engineering Physics, Tulane University, New Orleans, Louisiana 70118, USA*

Corresponding email address: yubo.drzhang@mju.edu.cn; jsun@tulane.edu

November 29, 2024

Transition-metal compounds (TMCs) with open-shell *d*-electrons are characterized by a complex interplay of lattice, charge, orbital, and spin degrees of freedom, giving rise to a diverse range of fascinating applications. Often exhibiting exotic properties, these compounds are commonly classified as correlated systems due to strong inter-electronic interactions called Hubbard *U*. This inherent complexity presents significant challenges to Kohn-Sham density functional theory (KS-DFT), the most widely used electronic structure method in condensed matter physics and materials science. While KS-DFT is, in principle, exact for the ground-state total energy, its exchange-correlation energy must be approximated in practice. The mean-field nature of KS implementations, combined with the limitations of current exchange-correlation density functional approximations, has led to the perception that DFT is inadequate for correlated systems, particularly TMCs. Consequently, a common workaround involves augmenting DFT with an on-site Hubbard-like *U* correction. In recent years, the *strongly constrained and appropriately normed* (SCAN) density functional, along with its refined variant r²SCAN, has achieved remarkable progress in accurately describing the structural, energetic, electronic, magnetic, and vibrational properties of TMCs, challenging the traditional perception of DFT's limitations. This review explores the design principles of SCAN and r²SCAN, highlights their key advancements in studying TMCs, explains the mechanisms driving these improvements, and addresses the remaining challenges in this evolving field.

## Contents

## 1. Introduction

  Transition-metal compounds (TMCs) exhibit a wealth of intriguing physics and are widely utilized in applications such as high-temperature superconductors, giant magnetoresistance, and catalysis. Their electronic properties span a broad spectrum, ranging from wide-gap insulators and narrow-gap semiconductors to metals, superconductors, and systems exhibiting metal-insulator transitions. Similarly, their magnetic properties are equally diverse, including ferromagnets with macroscopic magnetism and antiferromagnets with zero net magnetization. These fascinating properties and applications are fundamentally linked to the unique electronic structures of TMCs, which arise from the intricate interplay of four critical degrees of freedom: lattice, charge, orbital, and spin within the localized $d$ orbitals [1].

  The intricate interactions and competing effects among different degrees of freedom in TMCs pose significant challenges for accurate and reliable theoretical simulations, including Kohn-Sham density functional theory (KS-DFT) [2] which has long been the cornerstone of computational condensed matter physics and materials science. A widely recognized example of DFT's limitations was its inability to predict the insulating behavior of certain Mott insulators, such as the parent phase of cuprate superconductors. This shortcoming can be addressed by augmenting DFT with a Hubbard-like $U$ parameter [3], originally introduced in the Hubbard effective Hamiltonian to account for inter-electron repulsion. Over time, this limitation has contributed to the widespread belief that TMCs are inherently strongly correlated materials and that conventional DFT is inadequate for describing their properties [4].

  However, KS-DFT is an exact theory in principle, and it contains all the many-body quantum effects in the exchange-correlation (XC) energy functional. In practice, the XC density functional has to be approximated. It is a constant and vibrant effort to develop more accurate approximate functionals, and Perdew and his collaborator classified them into Jacob's ladder of DFT [5]. The most popular XC forms in the condensed matter and materials science communities have been the local density approximation (LDA; the first rung) [6], the generalized gradient approximation (GGA) in the form of Perdew-Burke-Ernzerhof (PBE; the second rung) [7], and the range-separated Heyd-Scuseria-Ernzerhof (HSE; the fourth rung) hybrid functional [8,9]. In 2015, Sun, Ruzsinszky, and Perdew proposed the *Strongly Constrained and Appropriately Normed* (SCAN) meta-GGA [10] on the third rung of Jacob's ladder. SCAN is fully constrained and obeys all 17 known exact constraints that a meta-GGA can, and it is also exact or nearly exact for a set of "appropriate norms" [10]. To enhance its numerical efficiency, J. Furness et al. proposed the restored-regularized SCAN ($r^2$SCAN) [11]. Since their inception, SCAN and $r^2$SCAN have garnered vast attention due to their significantly improved performance and small additional computational demands compared to LDA and PBE. Notably, $r^2$SCAN has been adopted by the Materials Project database since its release v2022.10.28 [12].

For TMCs, SCAN and $r^2$SCAN have also demonstrated remarkable performances, as evidenced by numerous evaluations [13-32] and a review [33]. For instance, SCAN reliably predicts the insulating phase of pristine $La_2SrCuO_4$, as well as the metallic phase of hole-doped $La_{2-x}Sr_xCuO_4$ [14], contrasting sharply with conventional functionals: LDA and PBE fail to open the bandgap in the insulating phase, while HSE incorrectly identifies the hole-doped metal as an insulator [25].

The mini-review summarizes significant developments to facilitate the broader application of the SCAN and $r^2$SCAN methods in modeling TMCs, as well as the mechanisms under these improvements. Selected examples include simple binary oxides, ferroelectric and multiferroic materials, cuprate and nickelate superconductors, and Heusler thermoelectric materials. This broad coverage highlights the versatility and effectiveness of SCAN in addressing the complex electronic structures and interactions intrinsic to these varied materials. The paper is structured as follows: the **introduction** provides an overview; **Section 2** introduces the design principles of SCAN and $r^2$SCAN; **Section 3** discusses SCAN's improvements in canonical finite systems due to the *self-interaction error* (SIE) reduction and spin-symmetry-breaking for total energy descriptions; **Section 4** demonstrates the calculations of electronic and magnetic properties of many materials, where SCAN and $r^2$SCAN improve over the conventional functionals; **Section 5** focuses on energetic and structural properties; **Section 6** addresses miscellaneous issues including remaining challenges and future development directions. Due to length constraints, this paper selectively covers a few topics to provide a systematic overview, primarily incorporating results from our previous studies while also introducing some new findings unique to this work.

## 2. Development of the meta-GGAs

DFT is, in principle, exact for determining the ground-state electron density and energy by reducing the many-electron problem to one-electron solutions of the auxiliary Schrödinger-like KS equations [2,34,35]. However, in practice, the exchange-correlation (XC) energy, which appears in the KS equations as a functional of the electron density, must be approximated. The XC energy can be expressed as:

$$E_{XC}[n_\uparrow, n_\downarrow] = \int d^3 r \, n(r) \epsilon_{XC}([n_\uparrow, n_\downarrow]; r) , \tag{1}$$

where $\epsilon_{XC}([n_\uparrow, n_\downarrow]; r)$ is the position-dependent XC energy per electron and is commonly approximated as $\epsilon_{XC}(n_\uparrow, n_\downarrow, \nabla n_\uparrow, \nabla n_\downarrow, \nabla^2 n_\uparrow, \nabla^2 n_\downarrow, \tau_\uparrow, \tau_\downarrow)$, depending only on local or semilocal information from the electron density. The local spin-densities are the only ingredients used in the local spin density approximations (LSDA) [36-38]. Spin-density gradients, $\nabla n_\sigma$, are added in GGAs [7,39-43]. The positive orbital kinetic energy densities $\tau_\sigma$ and/or the Laplacian of the electron density, $\nabla^2 n_\sigma$, are further added in meta-GGAs [10,44-50]. $\tau_\sigma$ dependent meta-GGAs are more prevalent and form the primary focus of this review. Nonlocal density functional approximations (DFAs) are constructed by including nonlocal ingredients, such as in hybrid density functionals that combine a semilocal functional with exact exchange energy or orbital-dependent functionals like the *Perdew-Zunger self-interaction correction* (PZ-SIC) [51,52].

There are two principal methodologies for developing DFAs [53]. The empirical approach determines parameters by fitting a proposed approximation to carefully curated experimental or high-level theoretical datasets. Conversely, the nonempirical approach avoids such datasets, relying on the mathematical properties and constraints of the exact functional. The latter approach maximizes the transferability of the functional across various systems and properties. SCAN and $r^2$SCAN fall under the meta-GGA category and were developed using the nonempirical approach, adhering to exact constraints to ensure broad applicability and reliability. There are many meta-GGAs [10,11,44,46,48-50,54-80], including those developed non-empirically [10,11,44,46,50,67-74], empirically [49,75-78], or by machine learning [79,80].

As more ingredients are incorporated, semilocal functionals can be designed to satisfy an increasing number of exact constraints with more appropriate norms. Unlike GGA functionals, meta-GGAs do not have to choose between exact constraints [10,44,45,81,82], as the additional flexibility provided by the kinetic energy density, $\tau$, allows all 17 exact constraints suitable for semilocal functionals to be satisfied within a single framework [10]. There are two formal reasons for the inclusion of $\tau$. First, it naturally arises in the Taylor expansion of the exact spherically averaged exchange hole near the reference point [79]. Second, using $\tau$ provides a simple and straightforward way to construct a correlation functional that is exactly one-electron self-interaction free [83]. Additionally, including the kinetic energy density also enables meta-GGAs to have the flexibility to satisfy more exact constraints and thus circumvent the "structure or energy" dilemma experienced by GGAs. By using a dimensionless variable $\alpha = (\tau - \tau^w)/\tau^{unif}$, where $\tau^{unif} = (3/10)(3\pi^2)^{2/3}n^{5/3}$ is the kinetic energy density of the uniform electron gas and $\tau^w = |\nabla n|^2/8n$ is the von Weizsäcker kinetic energy density that is exact for single orbital systems, meta-GGAs can recognize the slowly-varying densities ($\alpha \approx 1$, characterizing metallic bonds), the single-orbital systems ($\alpha = 0$, characterizing covalent single bonds), and noncovalent bonds with $\alpha \gg 1$ between two closed shells [84,85]. For the noncovalent bonds, e.g., the van der Waals bond of the Ar dimer around equilibrium, $\tau^w$ is zero at the bond center by symmetry, $\tau^{unif}$ scales as $n^{5/3}$ which is small, and $\tau$ is large due to the occupation of antibonding molecular orbitals. Thus, $\alpha$ is much greater than 1 for noncovalent bonds. $\alpha$ is directly related to the electron localization function (ELF) with $1/(1+\alpha^2)$, and therefore identifies different chemical bonds [82,86]. Despite the inclusion of $\tau$, the computational cost of meta-GGAs remains comparable to that of GGAs due to the semilocal nature of $\tau$. However, as $\tau$ is determined by KS orbitals, which are nonlocal functionals of the electron density, meta-GGAs are inherently nonlocal. Based on this property, Nazarov and Vignale demonstrated that the XC kernel derived from meta-GGAs within the adiabatic time-dependent DFT (TDDFT) framework introduces nonlocality — a feature absent in GGAs but essential for accurately describing excitonic effects in crystals [87].

By taking the above advantages, the nonempirical SCAN meta-GGA was recently developed [10]. SCAN is the first meta-GGA that is fully constrained, obeying all 17 known exact constraints that a semilocal functional can. It is also exact or nearly exact for a set of "appropriate norms", including rare-gas atoms and nonbonded interactions. SCAN recognizes different chemical bonds via $\alpha$ and treats them appropriately with different GGAs. SCAN thus predicts accurate geometries and energies of diversely-bonded molecules and materials (including covalent, metallic, ionic, hydrogen, and van der Waals bonds), significantly and systematically improving at comparable efficiency over its predecessors, the GGAs. The SCAN meta-GGA functional is constructed as,

$$\epsilon_{XC}^{SCAN} = \epsilon_X^{SCAN} + \epsilon_C^{SCAN} , \qquad (2)$$

where,

$$\epsilon_{X/C}^{SCAN} = \epsilon_{X/c}^{SCAN1} + f_{X/C}(\alpha)[\epsilon_{X/C}^{SCAN0} - \epsilon_{X/c}^{SCAN1}] . \qquad (3)$$

The energy densities per electron, $\epsilon_{X/C}^{SCAN0}$, are GGAs specifically designed to accurately describe the exchange or correlation energy per electron in single-orbital systems near equilibrium [10,88], where $\alpha = 0$. The $\epsilon_{X/c}^{SCAN1}$ GGAs are tailored for slowly varying densities, characterized by $\alpha = 1$. An interpolation was constructed to transition between the $\epsilon_{X/c}^{SCAN0}$ and $\epsilon_{X/c}^{SCAN1}$ GGAs as $f_{X/C}(\alpha)$, for the range $f_{X/C}(\alpha = 0) = 1$ and $f_{X/C}(\alpha = 1) = 0$, and extrapolated to $\alpha \gg 1$. An infinite number of functionals could be built from this form, and all would satisfy the 17 exact constraints. SCAN employs a set of appropriate norms to guide the functional systematically and logically from one exact constraint to another. The appropriate norms for a semilocal functional like SCAN are systems in which the exact XC hole, $n_{XC}^{exact}(\boldsymbol{r},\boldsymbol{r'})$, remains localized near the reference electron at position $\boldsymbol{r}$. For instance, the hydrogen atom serves as one of these appropriate norms [10]. In semilocal functionals, however, an error cancellation is expected between the exchange and correlation components. This issue arises because the exact XC

Page 4 of 41

hole is deeper and more localized near the reference electron than the exact exchange hole alone [10]. Semilocal density functionals leverage this error cancellation to provide a more accurate model of the exact XC functional as a whole.

The SCAN functional, while highly successful, has well-documented numerical issues [89,90] that limit its applicability to highly sensitive or large-scale problems. Additionally, the SCAN functional exhibits divergences in single-orbital systems, complicating the generation of pseudopotentials for hydrogen and helium atoms [91,92]. Modifications to SCAN, introduced by Mezei, Csonka, and Kállay, enhance the accuracy of atomization energies for multi-reference systems but fail to resolve its numerical sensitivities [93].

In response to this issue, Bartók and Yates proposed a regularization of SCAN termed "rSCAN" [91]. This modification made two substitutions. The first regularization is to the $\alpha$ iso-orbital indicator,

$$\alpha \to \alpha' = \frac{\tilde{\alpha}^3}{\tilde{\alpha}^2 + \alpha_r} , \tag{2}$$

where

$$\tilde{\alpha} = \frac{\tau - \tau^w}{\tau^{unif} + \tau_r} , \tag{3}$$

where $\alpha_r = 10^{-3}$ and $\tau_r = 10^{-4}$ are regularization constants. This regularization controls the single-orbital diverge at the expense of breaking the uniform density limit and coordinate scaling constraints obeyed by SCAN. The second regularization is to change the interpolation from the piecewise exponential function of SCAN, which introduces a kink at $\alpha \approx 1$, into a smooth polynomial interpolation (Figure 1).

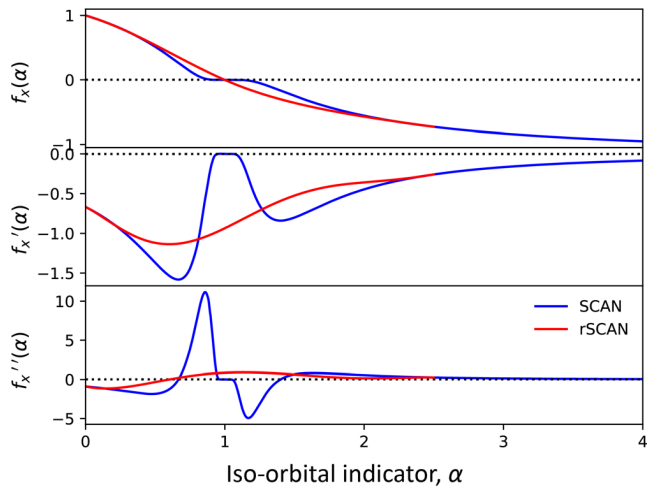

**Figure 1**. **Switching functions and their first and second derivatives used in SCAN and rSCAN.** The switching function, $f_x(\alpha)$, was introduced in the original SCAN functional in Ref. [10]. The figure is reproduced from Ref. [91].

While these modifications successfully resolve the numerical problems, they break four of the exact constraints obeyed by SCAN. As a result, the functional accuracy is less transferable and is degraded for atomization energies [91,94]. In subsequent work, Furness et al. modified rSCAN to restore adherence to the exact constraints while retaining its beneficial regularizations, resulting in the *restored-regularized SCAN* (r$^2$SCAN) functional [11]. r$^2$SCAN satisfies all the exact constraints of SCAN, except for the fourth-order gradient expansion for slowly varying densities. This restoration of exact constraints, combined with rSCAN's regularizations, enables r$^2$SCAN to match SCAN's accuracy while avoiding its numerical issues [11,94].

## 3. Understanding SCAN's performance in canonical molecules

Most popular DFAs encounter two key theoretical challenges: the *self-interaction error* (SIE) [51,95] and the *static correlation error* [96,97]. These two errors can lead to significant inaccuracies in predicting material properties, particularly for TMCs, exemplified by the underestimation of bandgaps in insulating materials [98]. We use canonical molecular examples to demonstrate how the SCAN/r²SCAN functional effectively addresses these issues, offering notable improvements over conventional functionals.

### 3.1. Self-interaction error and its reduction in SCAN

SIE arises from the incomplete cancellation of the spurious classical Coulomb self-interaction by the self-interaction XC components of DFAs [51,95]. Since the repulsive self-Coulomb term typically exceeds the attractive self-exchange-correlation term, the net SIE is typically positive. This results in orbitals being under-bound (orbital energies too high) and wavefunctions overly delocalized, leading to what is often termed the *delocalization erro*r. SIE is most clearly defined and illustrated in single-electron systems, where it should ideally be absent, as only one electron is present. The behavior of SIE in many-electron systems will be discussed later. Figure 2a shows the binding energy curve of the $H_2^+$ molecule calculated using the Hartree-Fock approximation, PBE, and SCAN functionals. The Hartree-Fock method, being orbital-dependent, serves as a reliable reference since its self-interaction free. Most widely used density functionals have not explicitly considered SIE into their design. Nevertheless, it is still possible to assess the extent of SIE reduction achieved by a functional, even if it was not explicitly designed to include this feature. As shown in Figure 2a, PBE and SCAN develop significant SIE when $H_2^+$ is stretched, with SCAN showing better SIE reduction than PBE.

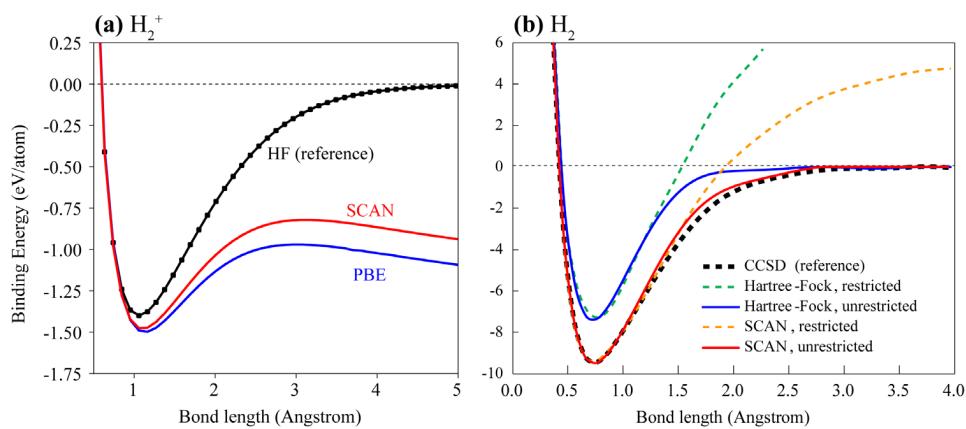

**Figure 2. Demonstration of self-interaction error and static-correlation error in DFT.** (a) Binding curve of a hypothetical $H_2^+$ molecule for demonstrating the self-interaction error. Hartree-Fock provides reference results. (b) Binding curve of an $H_2$ molecule for displaying the static-correlation error. The many-body CCSD (Coupled Cluster Singles and Doubles) results are taken as a reference. Compiled from Ref. [15].

### 3.2. Static correlation and the spin-symmetry-breaking technique

Figure 2b presents the binding energy curve of the $H_2$, highlighting the emergence of static correlation issues as the molecule is stretched. This stretching leads to a degeneracy between the first excited spin-triplet state and the ground-state spin-singlet state. In such cases, static or non-dynamic correlation arises due to the degeneracy or near-degeneracy of the ground-state Slater determinant and multireference effects. The coupled cluster method with singles and doubles (CCSD) for $H_2$ is exact within the given basis set, providing invaluable insights into its electronic structure. While exact solutions for multi-electron systems typically require exponentially scaling methods like full configuration-interaction (FCI), the small size of the two-electron $H_2$ system allows for exact diagonalization, making CCSD equivalent to FCI in this case, though it is generally unreliable for strongly correlated systems with more than two electrons.

The Hartree-Fock method offers a robust zero-order approximation, primarily capturing dynamical correlation effects. However, when spin symmetry is restricted, Hartree-Fock calculations result in overly positive binding energies compared to the CCSD reference, revealing a significant static correlation error inherent in the method. Allowing *spin-symmetry-breaking* (SSB) provides a partial remedy to improve the description of correlation energy. A spin-unrestricted Hartree-Fock model can achieve the exact dissociation limit, although it deviates from the reference curve at shorter bond distances.

SSB is equally critical in the context of DFT, as recently discussed by Perdew et al. [4,99]. Their work emphasized that "*spin-symmetry breaking can reveal strong correlations among electrons present in a symmetry-unbroken wavefunction*" and that SSB can transform strongly correlated systems into normally correlated systems, which approximate functionals can better handle. As shown in Figure 2b, the SCAN functional performs better than the Hartree-Fock method without SSB, although static correlation errors persist. Notably, when combined with SSB, the SCAN functional closely matches the entire reference binding curve, with discrepancies of at most approximately 0.1 eV/atom around 1.5 Å. This high level of accuracy is achieved by employing spin-contaminated unrestricted KS wavefunctions, which are no longer eigenfunctions of the total spin operator, $\hat{S}^2$ [100]. In subsequent sections, the SSB approach will be used unless stated otherwise.

### 3.3. More realistic bandgap within the generalized Kohn-Sham scheme

The term "*gap*" typically has two distinct definitions depending on the context in condensed matter physics and materials science: *fundamental energy gap* ($G$) and *single-particle bandgap* ($g$). The former $G$ is defined as the separation between the ionization energy ($E_I$) and the electron affinity ($E_A$), expressed as $G = E_I - E_A$. Here, the electron affinity is given by $E_A = E(N) - E(N+1)$, where $N$ represents the number of electrons. Since this definition involves only ground-state energies, $G$ can in principle be calculated exactly within DFT and directly compared to experimental measurements [98,101,102]. The latter $g$ is defined as $g = \varepsilon^{CBM} - \varepsilon^{VBM}$, where $\varepsilon^{CBM}$ and $\varepsilon^{VBM}$ are the conduction band minimum and valence band maximum, respectively. These values are derived from a KS-DFT or a generalized KS-DFT (gKS-DFT) calculation. In the KS-DFT framework, the XC potential is multiplicative, and the exact KS potential exhibits a derivative discontinuity when an electron is added to a neutral solid. However, this discontinuity is absent in LDA and GGAs, leading to the well-known underestimation of the bandgap. Correcting the KS bandgap by adding this discontinuity results in an accurate prediction of the fundamental gap.

Unlike the KS-DFT scheme, the gKS-DFT scheme allows the effective potential to be non-multiplicative and continuous, such as in orbital-dependent approximations. It has been proven that within the gKS-DFT framework, the single-particle bandgap ($g$) calculated using a density functional is equal to the fundamental energy gap ($G$) for the same functional, provided the gKS-DFT potential operator is continuous and the density change is delocalized when an electron or hole is added [98,101,102]. This result implies that improved density functionals designed for better total energies can also provide more accurate bandgaps for solids within the gKS-DFT scheme.

**Table 1. Energies and gaps of finite linear H$_2$ chain, extrapolated to infinite limit.** Ionization energy $E_I$, electron affinity $E_A$, and fundamental energy gap $G = E_I - E_A$; The band edge energies $\varepsilon^{VBM}$, $\varepsilon^{CBM}$, and single-particle bandgap $g = \varepsilon^{CBM} - \varepsilon^{VBM}$. The calculation uses the LDA, PBE GGA, SCAN meta-GGA, and HSE06 range-separated hybrid functionals. All results are in units of eV. Cited from Ref. [98].

|       | $(E_I+E_A)/2$ | $E_I$ | $-\varepsilon^{VBM}$ | $E_A$ | $-\varepsilon^{CBM}$ | $G$ | $g$ |
|---|---|---|---|---|---|---|---|
| LDA   | 1.65 | 3.14 | 3.13 | 0.16  | 0.17  | 2.98 | 2.96 |
| PBE   | 1.67 | 3.24 | 3.23 | 0.09  | 0.10  | 3.15 | 3.13 |
| SCAN  | 1.68 | 3.33 | 3.31 | 0.01  | 0.02  | 3.32 | 3.29 |
| HSE06 | 1.82 | 3.92 | 3.91 | −0.29 | −0.28 | 4.21 | 4.18 |

SCAN, being orbital-dependent and implemented within the gKS-DFT scheme with a continuous effective potential, predicts more realistic bandgaps than those from GGAs or even the exact KS-DFT potential [98]. To illustrate this, Perdew et al. [98] calculated the bandgap of a finite one-dimensional linear chain of H$_2$ molecules and extrapolated the results to the infinite limit, as shown in Table 1. The results demonstrate, within numerical accuracy, that, $E_I \to -\varepsilon^{VBM}$, $E_A \to -\varepsilon^{CBM}$, and $G \to g$. Furthermore, the bandgaps calculated using SCAN and HSE06 functionals (both implemented in the gKS-DFT scheme) are larger than those obtained with LDA and PBE, which operate in the KS-DFT scheme. It should be noted, however, that the $H_2$ chain model is not a realistic system, and its experimental gap is unknown.

## 4. Electronic and magnetic properties

Given the lattice, charge, orbital, and spin degrees of freedom in TMCs, we simplify our analysis by categorizing these materials into three groups with increasing complexity. Section 4.1 addresses structurally simple semiconductors, focusing on their electronic structures and the calculation of bandgaps in narrow-gap materials. Section 4.2 explores functional materials with complex crystal structures, highlighting essential physics within unitcell models. Section 4.3 delves into highly complex materials characterized by nanoscale spatial inhomogeneities, where DFT using cost-effective functionals remains the only viable first-principles method.

### 4.1. Semiconductors with simple and high-symmetric crystal structures

This subsection starts with band structure calculations for three semiconductors, each with high symmetry and no internal structural freedom. Elemental Ge exhibits simple $sp$ bonding, but its narrow bandgap poses computational challenges. Zinc-blende ZnS, with the $3d$ orbitals distant from the band edge, makes its DFT calculation relatively straightforward, serving as a reference system. In contrast, antifluorite $Cu_2S$ presents significant challenges in band structure calculation due to its narrow bandgap and shallow $3d$ orbitals. Subsequently, we explore the band structures of four transition-metal monoxides (i.e., MnO, NiO, FeO, and CoO), of which conventional DFAs often struggle to open their bandgaps.

#### 4.1.1. Full-yet-shallow 3d shells in chemical bonding

A narrow bandgap presents a significant computational challenge to common DFAs, even without the involvement of localized $d$ electrons. Diamond-structure Ge (Figure 3a), which has four $sp^3$ valence electrons, exhibits a narrow bandgap of 0.74 eV at zero temperature [103]. Despite its electronic simplicity, both LDA and PBE functionals fail to open the bandgap, instead predicting a semimetal with topologically inverted band ordering. As illustrated in Figure 3d, the expected Ge-$5s$ conduction states at the Brillouin zone center are incorrectly inverted into the valence bands, resulting in a negative bandgap of −0.04 eV. In contrast, r$^2$SCAN predicts a positive bandgap of 0.385 eV (Figure 3e), although smaller than the experimental values. It is worth noting that the r$^2$SCAN bandgap improves over a previous SCAN calculation of Ge [89].

The zincblende structure of ZnS can be derived from Ge through atomic substitution while reserving the valent electron counts (Figure 3b): $2Ge^{4e} \to Zn^{2e} + S^{6e}$. Since the Zn-$d^{10}$ shells are fully occupied and away from the valence-band edge, it induces marginal theoretical complications. The PBE bandgap is 2.10 eV (Figure 3f), strongly underestimated compared to the experimental value of 3.7 eV. The r$^2$SCAN bandgap increases to 2.66 eV, and the Zn-$3d$ energetic levels shift lower by approximately 0.2 eV (Figure 3g) compared to the PBE results.

Further cation substitution in ZnS leads to the formation of antifluorite $Cu_2S$ (Figure 3c): $Zn^{2e} \to 2Cu^{1e}$. In $Cu_2S$, the fully occupied Cu-$3d$ shells are so shallow that they actively participate in bonding interactions with the S-$3p$ orbitals [104]. Experimentally, $Cu_2S$ is identified as a narrow-gap (0.75 eV) semiconductor used as photovoltaic and thermoelectric material [104]. However, PBE incorrectly predicts a topologically nontrivial semimetal, with significant band inversion at the Brillouin zone center (−0.92 eV; Figure 3h). The absence of a bandgap is due to the Cu-$3d$ orbitals being calculated too high by PBE, which exacerbates $pd$ repulsion and closes the bandgap [104]. While the incorrect band inversion still occurs in the r$^2$SCAN simulation, its magnitude is reduced to 0.50 eV (Figure 3i). The Cu-$3d$ states also shift downward in r$^2$SCAN.

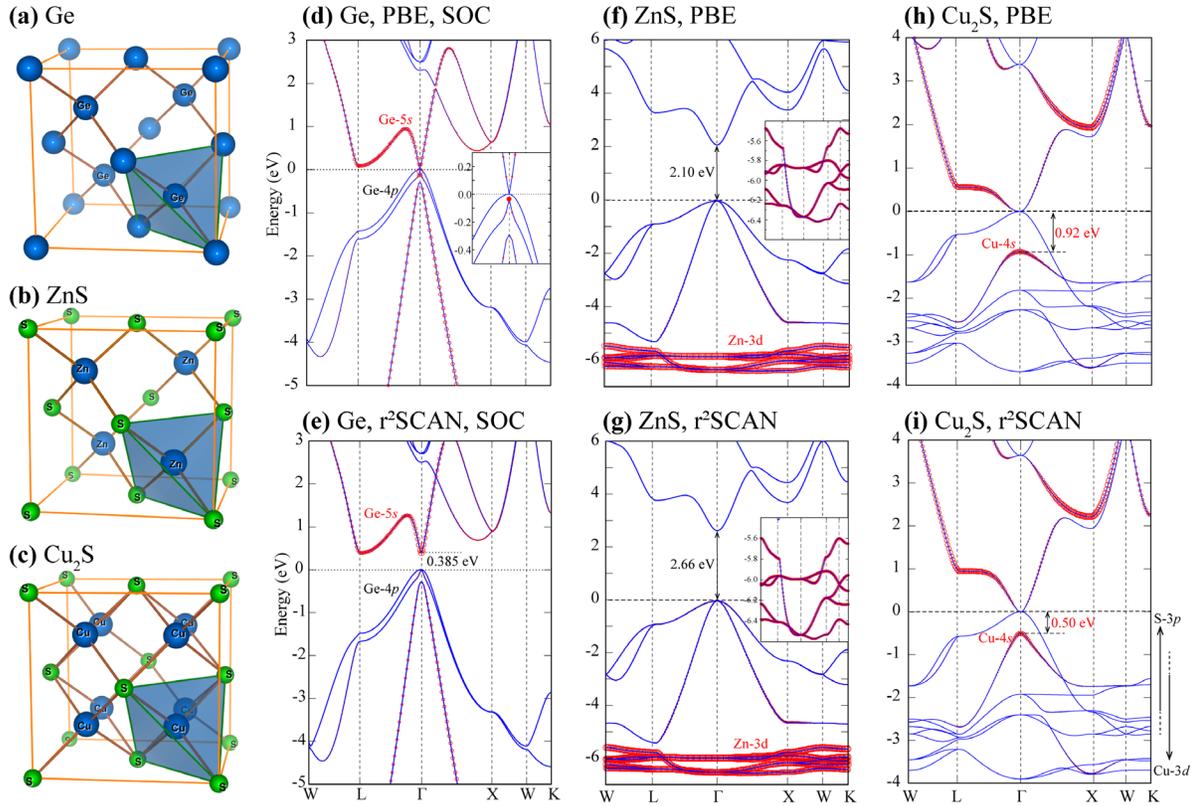

**Figure 3. Crystal structures and band structures of simple semiconductors.** (a,b,c) Crystal structures of the diamond structure Ge, zincblende ZnS, and antifluorite $Cu_2S$. (d-i) Band structures calculated by PBE and $r^2$SCAN. The red circles represent the wavefunctions of Ge-5$s$, Zn-3$d$, and Cu-4$s$ orbitals. The insets are the zoom-in views of the band structures. Spin-orbit coupling is considered in calculating Ge's band structure.

How does $r^2$SCAN improve band structure calculations compared to PBE? We have demonstrated that the significant SIE reduction is one of the primary improvements of the SCAN functional, as applied to the canonical $H_2^+$ molecule. We have also shown SCAN's bandgap improvement in the linear $H_2$ chain. For solids, we here present an intuitive understanding by plotting the electron redistribution, $\Delta n = n^{r2SCAN} - n^{PBE}$, in Figure 4.

(1) For Ge (Figure 4a), there is a noticeable electron depletion at the Ge sites and an accumulation at the bond centers. This electron redistribution enhances covalent bonding in the $r^2$SCAN simulation, effectively addressing the under-binding issue commonly noted with PBE [105]. The strengthened covalency in $r^2$SCAN leads to more realistic band-edge energies, which restores the correct band order (Figure 3e).

(2) In ZnS, the electron redistribution patterns separate into several areas. The electron accumulation at the inner area of the Zn sites (labeled as 1 in Figure 4b) suggests that Zn-3$d$ orbitals become more localized and compact, which correlates with the downward shift of the 3$d$ bands (inset of Figure 3g). Conversely, the outer regions (labeled as 2) show electron depletion, indicating that the 4$s$ electrons become more delocalized or are transferred away from these areas. The pattern around the S sites is particularly intricate: electron accumulation (area-1) and depletion (area-2) suggest localization of the S-3$s$ and 3$p$ orbital; area-3 exhibits electron accumulation, indicating electron transfer from the neighboring Zn 4$s$ orbitals. This observation of *inter-site electron transfer* is supported by ionic valence (Figure 4d): Zn's valence increases from +0.985 in PBE to +1.020 in $r^2$SCAN. Overall, the electron redistribution contributes to $r^2$SCAN's larger bandgap in two ways: the more ionic Zn-S bond and the reduced $pd$ repulsion associated with the deeper Zn-3$d$ orbitals [106].

(3) Similar results are observed in Cu$_2$S (Figure 4c), underscoring the consistency and effectiveness of the r$^2$SCAN approach in capturing these details of the electronic structure. However, the r$^2$SCAN band structure still has band inversion, which is particularly puzzling given that the Cu-3$d$ orbitals are fully occupied. The problem might originate from the specific 3$d$ orbital delocalization in Cu$_2$S, as addressed in references [104]. To simultaneously reduce the delocalization error and accurately predict the bandgap, the mBJ potential [107] and TASK functional [71] were combined with a Hubbard-like $U$ correction, which better reproduces the correct band topology and bandgap values [108].

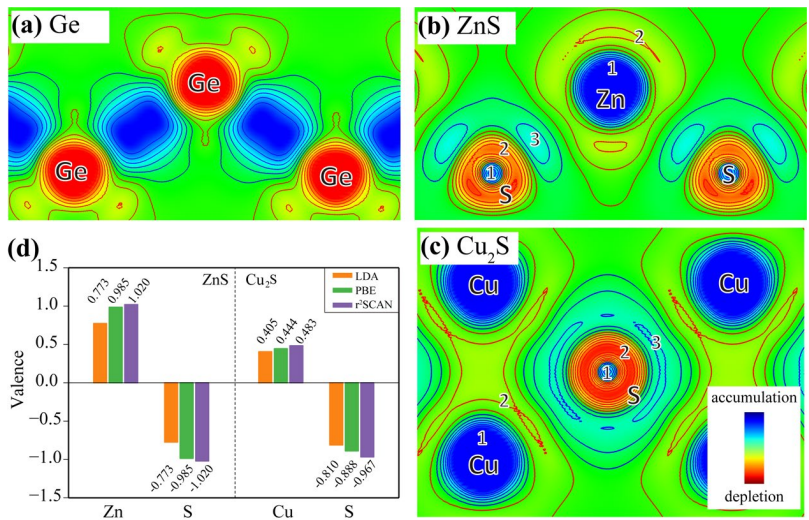

**Figure 4. Modulations of electron density and ionic valence.** (a,b,c) Electron redistribution from PBE to r$^2$SCAN, $\Delta n = n^{\text{r2SCAN}} - n^{\text{PBE}}$, in Ge, ZnS, and Cu$_2$S. The red and blue contours indicate electron depletion and accumulation, respectively. (d) Valence states of ZnS and Cu$_2$S.

### 4.1.2. Monoxides with open 3$d$-shells: bandgap opening without the Hubbard-like $U$ parameter

Transition-metal monoxides such as MnO, NiO, FeO, and CoO, which possess open 3$d$ shells, are commonly classified as Mott insulators (may mix with charge transfer characteristics). When DFT fails to open the bandgaps, it is then cited as DFT's inability to describe the (strong) electron correlation. As depicted in Figure 5a, while PBE opens bandgaps for MnO and NiO, it inaccurately predicts that FeO and CoO are metallic. A widely adopted remedy is to augment DFT with a Hubbard-like $U$ potential derived from the Hubbard effective Hamiltonian to address inter-electron repulsion [3]. As demonstrated in Figure 5b, the PBE+$U$ approach opens the bandgaps of FeO and CoO. The success of the +$U$ approach has reinforced the perception that a strong correlation means "*everything DFT gets wrong*" [4], leading to the prevailing view that DFT should invariably be combined with the $U$ correction for TMCs.

Essentially, the DFT+$U$ approach (typically with a positive $U$ value) applies a corrective potential, $\Delta V = V_{\text{DFT}+U} - V_{\text{DFT}} = U(\frac{1}{2} - n_i)$, onto the $d$ orbitals [3]. $\Delta V$ is negative when an orbital is filled ($n_i = 1$) and positive when it is empty ($n_i = 0$); this splitting contributes to the opening of a bandgap. Since $\Delta V$ lowers the potential of the occupied states by an amount of $U/2$, the $U$ correction also improves the spatial localization of the $d$ orbitals [109]. From the perspective of fractional charge, bandgap underestimation in local and semi-local functionals is often attributed to the derivative discontinuity near integer electron numbers [101]. In this context, the $U$ parameter partially corrects this discontinuity, enhancing bandgap predictions.

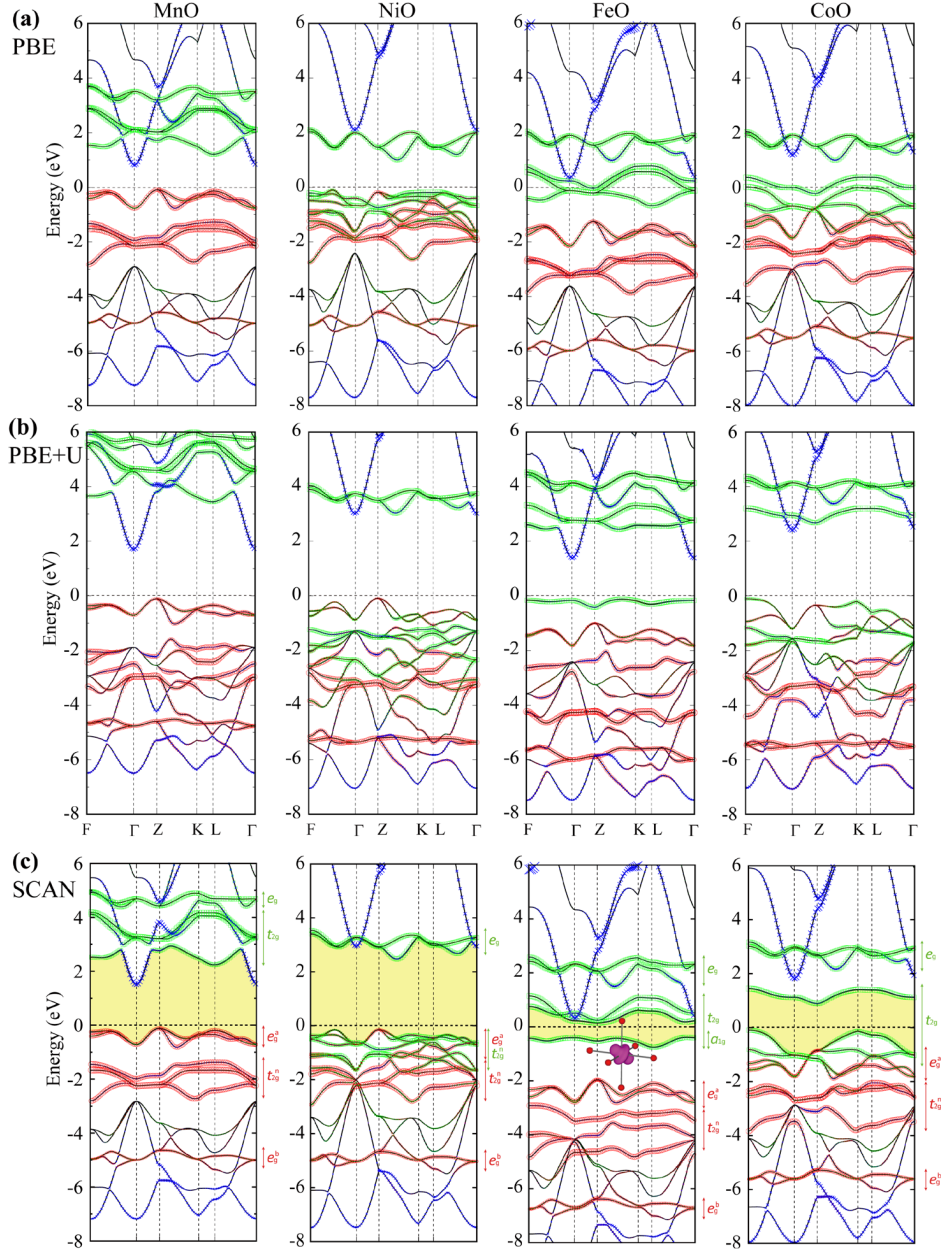

**Figure 5. Band structures of transition-metal monoxides.** The calculations use the experimental NaCl-type geometries and G-type antiferromagnetic spin configuration. The top panel (a), middle panel (b), and bottom panel (c) are the PBE, PBE+$U$, and SCAN results, respectively. Orbital characters are indicated as: red circles are the spin-majority 3$d$ states in one spin channel, green squares are the spin-minority 3$d$ states, and blue crosses are the transition-metal 4$s$ states. The $U$ values are 3.9 eV for MnO, 4.3 eV for NiO, 3.9 eV for FeO, and 3.2 eV for CoO. Subplot (c) is compiled from Ref. [15].

Interestingly, SCAN alone (without $U$) opens all the bandgaps (Figure 5c). This success demonstrates that accurate calculations of so-called correlated materials do not necessarily rely on the Hubbard-like $U$, affirming DFT's capability to capture the essential physics of these materials. Based on the SCAN band structure, the bandgap mechanisms were explored in reference [15] and are briefly reproduced here. (1) In MnO, the bandgap is due to exchange splitting between the spin-up and spin-down channels, resulting in the Mn$^{2+}$ 3$d^5$ orbitals having one fully occupied [$(t_{2g}^3 e_g^2)^\uparrow$] group and one empty [$(t_{2g}^0 e_g^0)^\downarrow$] group. (2) For NiO, the bandgap is between $(t_{2g}^3)^\downarrow$ and $(e_g^0)^\downarrow$ states, originating from NiO$_6$ octahedral crystal-field splitting.

(3) $Fe^{2+}$ with $d^6$ configuration in FeO has its majority spin channel filled by five electrons, leaving one electron in $t_{2g}$ orbitals in the minority spin channels. The $t_{2g}$ degeneracy must be lifted to open a bandgap across them: the splitting occurs between one occupied band and two empty bands. The occupied band is a linear combination of $d_{xy}$, $d_{yz}$, and $d_{xz}$ orbitals, demonstrating the *orbital occupation polarization*. (4) The bandgap opening mechanism of CoO is similar to FeO's, except that two electrons are polarized from three Co-$t_{2g}$ orbitals.

It is encouraging that SCAN, benefiting from the SIE-reduction, effectively captures all the bandgap mechanisms. It would be interesting to gain a more intuitive understanding of SCAN's specific performance in monoxides. In the below subsections, we categorize three effects of SIE-reduction: more compact orbitals (Section 4.1.2.1), effective inter-site electron redistribution (Section 4.1.2.2), and enhanced capture of orbital anisotropy (Section 4.1.2.3). Before delving into details, we briefly discuss the SSB method used in the solids. All four monoxides exhibit local magnetic moments arranged into G-type antiferromagnetic configurations at low temperatures, and fixing the static spin order in simulations breaks the spin symmetry. While the SSB method might be considered more of a theoretical assumption in finite molecules (such as the stretched $H_2$ in Figure 2b), it "*may correspond to an actual state*" in the extended solids, as noted by Martin, Reining, and Ceperley [110]. Philip Anderson [111] stated that the span for spin flips in the antiferromagnetic NiO is typically three years, far exceeding the duration of a neutron scattering experiment [4]. Furthermore, Perdew et al. [4] recently highlighted that SSB "*can reveal strong correlations among electrons present in a symmetry-unbroken wavefunction*" if the broken symmetry persists over a long period. Conversely, conserving spin symmetry leads DFAs to exhibit an explicitly discontinuous (non-differentiable) dependence on density or density matrix. This scenario, known as the non-magnetic phase in monoxides, results in excessively high energies in the simulation and is not reflective of the experimental state [15].

### 4.1.2.1. Consequence 1 of SIE-reduction: More compact *d* orbitals

In the case of many-electron systems, such as monoxides, SIE is characterized by deviations from the ideal piecewise linearity of total energy between two consecutive integer electron counts, as demonstrated by Mn ions in Figures 6a and 6b. Ideally, an exact density functional would maintain a linear relationship between these integer counts, with clear derivative discontinuities at each integer [112], thus meeting the generalized Koopman's condition. Many commonly used local and semi-local density functionals (e.g., PBE and SCAN) tend to stabilize the energy excessively for fractional electron numbers, leading to a concave energy curve. This deviation from linearity suggests a tendency to spuriously delocalize electrons across nuclear centers, a phenomenon recognized as the *delocalization error*. Figures 6a and 6b demonstrate that the delocalization error continues to affect SCAN, but it is effectively mitigated compared to that in PBE. On the other hand, the Hartree-Fock approximation, which is orbital-dependent and exact for single-electron systems, thereby avoiding SIE, adheres to Koopman's theorem. However, Hartree-Fock overlooks correlation energy and effective exchange interaction screening, resulting in substantial overestimations of bandgaps. The resulting energy curve is convex, indicative of the *over-localization error*. Consequently, while orbital dependence is essential for a density functional, adherence to the generalized Koopman's condition is equally crucial for accurately predicting total energies and bandgaps using orbital energies.

Correcting one-electron SIE based solely on vacuum power-law behavior is insufficient for improving bandgap predictions. Reducing SIE also leads to more compact orbitals, effectively decreasing the delocalization error. In DFT calculations using the gKS scheme, the many-electron SIE predominantly influences the accuracy of bandgap predictions from orbital-dependent density functionals. Lower many-electron SIE corresponds to reduced errors in these predictions [97,101,113]. Several density functionals that are free from one-electron SIE, including weighted density approximation [114] and the PZ-SIC [51], have also demonstrated reduced many-electron SIE and enhanced bandgap predictions [113,115-118]. Notably, SCAN opens the bandgaps of FeO and CoO, and improves predictions for MnO and NiO by differentiating between occupied and unoccupied states and promoting more compact orbital formations (Figure 6c). From an energetic perspective, a more compact orbital typically possesses a lower eigenvalue, positioning it further from the valence band edge and contributing to a larger bandgap.

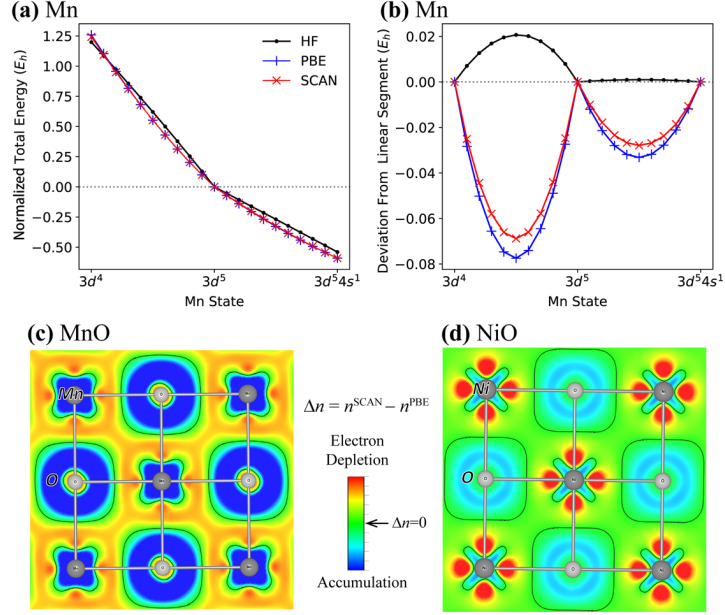

**Figure 6. Many-electron self-interaction error and electron density.** (a) The total energy of an isolated Mn ion as a function of orbital occupation, normalized such that the Mn$^{2+}$ $3d^5$ state is at zero. (b) Deviation from ideal linear behavior between integer electron numbers for the total energies. (c) Difference of electron density $\Delta n = n^{\text{SCAN}} - n^{\text{PBE}}$. The red and blue colors indicate electron depletion and accumulation, respectively, with their boundary of $\Delta n = 0$ delineated by black lines. The primary change is the increasing compactness of both Mn and O orbitals. (d) $\Delta n$ for NiO. In addition to the shrinkage of Ni and O orbitals, there is also significant electron depletion from Ni sites. Subplots (a) and (b) are compiled from Ref. [15].

### 4.1.2.2. Consequence 2 of SIE-reduction: Effective inter-site electron redistribution

The comparison between electron redistributions in MnO (Figure 6c) and NiO (Figure 6d) reveals an additional feature in NiO: significant electron depletion from Ni orbitals along the Ni-O bond direction when comparing the PBE and SCAN simulations. Analysis of the ionic valence, as shown in Figure 7c, indicates that these electrons are transferred from Ni to O sites. Similar behavior is also observed in FeO and CoO. By contrast, MnO exhibits almost identical valence states across the tested functionals. These observations highlight a second consequence of SIE reduction: *effective intersite electron redistribution* from PBE to SCAN calculations. The r$^2$SCAN results are similar to those from SCAN, and we omit separate discussions.

The significance of electron redistribution lies in its comprehensive impact on electronic structure. To understand these effects more thoroughly, we compare the performance of the SCAN method with the PBE+$U$ approach in calculating the bandgap, local magnetic moment, and cation valence, as shown in Figure 7. This comparison is crucial for appreciating the unique attributes of SCAN relative to the +$U$ correction, which has been well-documented [3,109]. As discussed earlier, the PBE+$U$ method modifies the bandgap (Figure 7a) by shifting the occupied *d*-orbitals downward and the unoccupied *d*-orbitals upward, primarily affecting the localized *d*-orbitals with minimal influence on the delocalized orbitals. Consequently, the *d*-orbitals become more spatially localized, increasing the magnetic moments (Figure 7b). In contrast, SCAN's enhancements extend beyond merely localizing *d*-orbitals; it also promotes more pronounced ionic bonding through inter-site electron redistribution (also refer to discussions Section 4.1.1). Regarding the bandgap formation, Figure 5c clarifies that while the near-edge valence bands predominantly consist of localized *d*-orbitals, the bottom conduction bands also incorporate delocalized *s*-states. Given its comprehensive enhancements across all orbital types, SCAN offers a more realistic description of the bandgap, capturing the underlying physics more comprehensively than the +$U$ method. Another more illustrative example can be found in Section 4.2.2, which discusses YMnO$_3$.

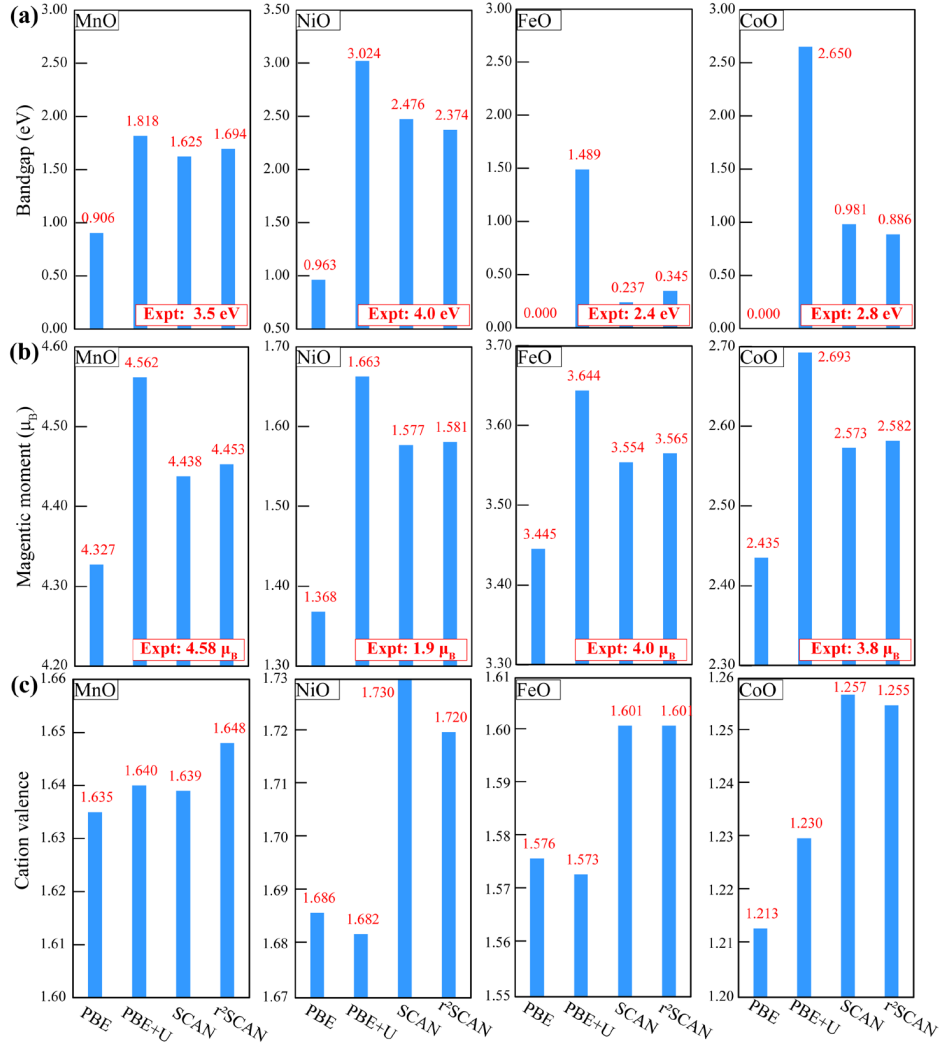

**Figure 7. Electronic and magnetic properties of transition-metal monoxides.** (a) Bandgap. (b) Local magnetic moment. (c) cation valence. The $U$ values are 3.9 eV for MnO, 4.3 eV for NiO, 3.9 eV for FeO, and 3.2 eV for CoO. The SCAN results are compiled from Ref. [15], while the other data are new from this work.

### 4.1.2.3. Consequence 3 of SIE-reduction: Enhanced capture of the orbital anisotropy

We have noted that bandgap formation in FeO and CoO relies on orbital polarization, where one or two $t_{2g}$ orbitals distinguish themselves from the other $t_{2g}$ orbitals. Such orbital anisotropy poses a significant challenge to DFAs, and SCAN's effectiveness in opening the bandgaps showcases its enhanced capabilities in this respect. Here, we provide more profound insights using FeO as an example. Figure 8a shows the PBE band structure, where the orbital symmetry is enforced to match the lattice symmetry—a common tactic to reduce computational costs. No method, including the PBE+$U$ and SCAN, can open a bandgap among the Γ-Z bands due to enforced $t_{2g}$-degeneracy. Upon lifting the orbital symmetry restriction, PBE creates small local gaps at each K-point (Figure 8b); however, these gaps are too narrow to form a global bandgap across the entire Brillouin zone. In contrast, SCAN effectively separates an occupied band (i.e., the $a_{1g}$ band) from the other unoccupied bands, leading to a discernible overall bandgap (Figure 8c).

The polarization of the $a_{1g}$ orbital, a linear combination of the three $t_{2g}$ orbitals [visualized using Wannier functions in Figures 8e to 8h], indicates intrinsic anisotropy among the $t_{2g}$ orbitals. According to SCAN calculations, the occupation numbers are $n(d_{xy}) = 0.332$, $n(d_{yz}) = 0.342$, and $n(d_{xz}) = 0.272$ electrons, highlighting the distinct nature of the $d_{xz}$ orbital compared to the similar $d_{xy}$ and $d_y$ orbitals. SCAN's ability to capture this orbital anisotropy is further demonstrated by the electron density analysis. In Figure 8i, the electron redistribution from calculations without symmetry to those preserving symmetry (i.e., $\Delta n^{(a)} = n^{\text{No-symmetry}} - n^{\text{Symmetry}}$) clearly illustrates the shape of the $a_{1g}$ orbital. A similar shape is also observed when comparing the differences between SCAN and PBE (i.e., $\Delta n^{(b)} = n^{\text{SCAN}} - n^{\text{PBE}}$), as shown in Figure 8j. The similarity between these two plots (i.e., $\Delta n^{(a)}$ and $\Delta n^{(b)}$) reveals that while PBE significantly underestimates the orbital anisotropy, SCAN more accurately captures it, consistent with SCAN's design principles of recognizing chemical environments and bonds.

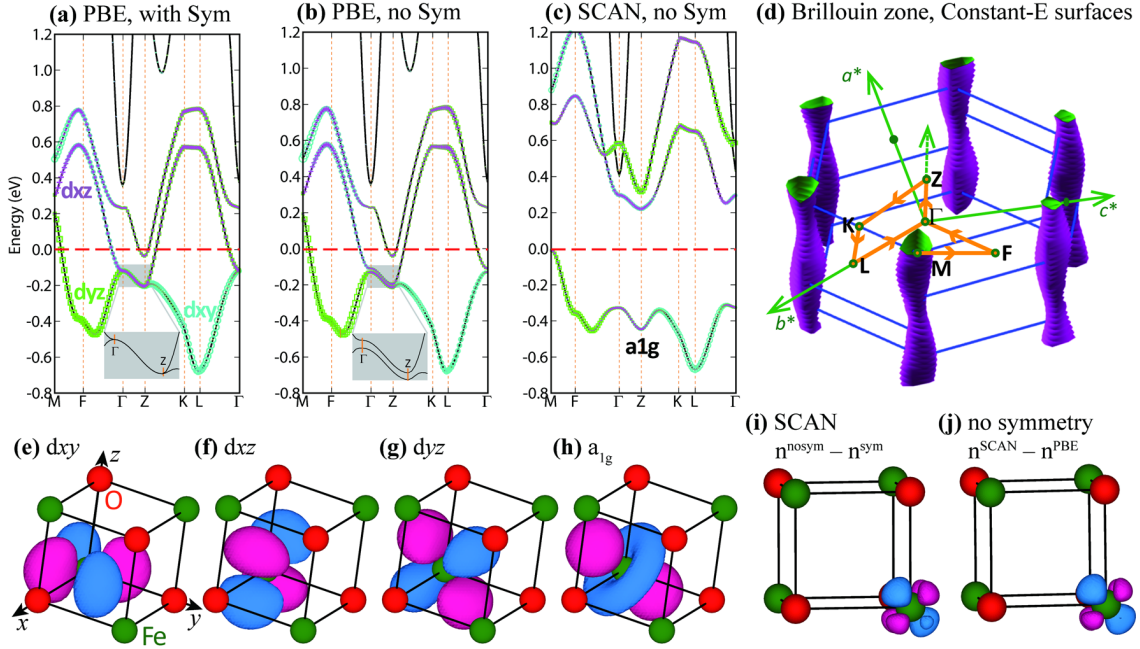

**Figure 8. Orbital polarization and anisotropy in FeO.** (a) PBE predicted band structure by keeping the $t_{2g}$ orbital symmetry. (b) PBE band structure but with the $t_{2g}$ orbital symmetry removed. (c) SCAN band structure with orbital symmetry removed. (d) Brillouin zone of the G-AFM phase, together with the constant energy surfaces that are 0.1 eV below the valence band maximum in subplot (c). (e-g) The $d_{xy}$, $d_{xz}$, and $d_{yz}$ orbitals from the Wannier function construction. The blue and pink colors denote the wavefunction signs. (h) The $a_{1g}$ orbital. (i) The difference of charge density distribution between calculations with and without orbital symmetry, both from SCAN calculations. The blue and pink colors denote charge accumulation and depletion, respectively. (j) The difference of charge density distribution, both without orbital symmetry. All calculations are for the cubic rock-salt structure, without geometrical distortions. Cited from Ref. [15].

### 4.1.3. Section summary

SCAN, being orbital-dependent and implemented within the gKS-DFT framework, accurately reproduces the near-edge band topology of the narrow-gap semiconductor Ge. In contrast, the PBE functional incorrectly predicts Ge as a semi-metal. Additionally, SCAN successfully opens the bandgap in transition-metal monoxides (MnO, NiO, FeO, and CoO) without relying on the Hubbard-like $U$ parameter. The enhanced performance of SCAN is attributed to the SIE-reduction, which manifests in three key ways: more compact $d$ orbitals, effective inter-site electron redistribution, and enhanced capture of orbital anisotropy. However, further efforts are required to accurately calculate the band structure of $Cu_2S$.

### 4.2. Ternary and quaternary oxides with complex geometries

Complex materials, often containing transition-metal elements and characterized by geometrical or electronic complexities, catalyze various intriguing functionalities. At the atomic level, these materials typically feature intricate and

Page 15 of 41

competing bond orders associated with *d*-orbitals, providing a rigorous test bed for assessing DFT's capabilities. In this section, we demonstrate the effectiveness of SCAN in simulating the electronic properties and functionalities of displacive ferroelectrics, magnetoelectric multiferroics, and cuprate high-temperature superconductors.

### 4.2.1. Displacive-ferroelectrics: Subtle bonding interactions due to geometry-symmetry-breaking

Displacive-type ferroelectric materials, such as $BaTiO_3$ and $PbTiO_3$ (Figure 9a), lose centrosymmetry upon cooling [119]. The distortion's strength stems from anisotropic bonding around Ti ions: While the elongated Ti-O bond exhibits increased ionic attributes, the opposite shortened bond has enhanced covalent properties. The ferroelectric behavior is highly dependent on the cation displacement, which is influenced by a delicate balance between the strengths of covalent and ionic bonds. The off-center atomic displacement leads to a tetragonal deformation characterized by $\eta = c/a > 1$. The calculated $\eta$ sensitively depends on the DFAs (Figure 9b): whereas $\eta$ is relatively well calculated by LDA, it is strongly overestimated by PBE, known as the *super-tetragonality* problem [120]. For instance, the PBE functional overestimates $\eta = 1.054$ for $BaTiO_3$ and $\eta = 1.240$ for $PbTiO_3$, compared to the experimental values of 1.010 and 1.071, respectively [31]. Spontaneous polarization ($P_s$) is one of the most critical properties of ferroelectrics, typically cited as 26 μC/cm² for $BaTiO_3$ and 57 μC/cm² for $PbTiO_3$ (Figure 9d). Using these values as benchmarks, the oldest LDA yields accurate $P_s$, whereas PBE significantly overestimates $P_s$. This scenario highlights the intriguing and somewhat puzzling superior performance of LDA in calculating $\eta$ and $P_s$.

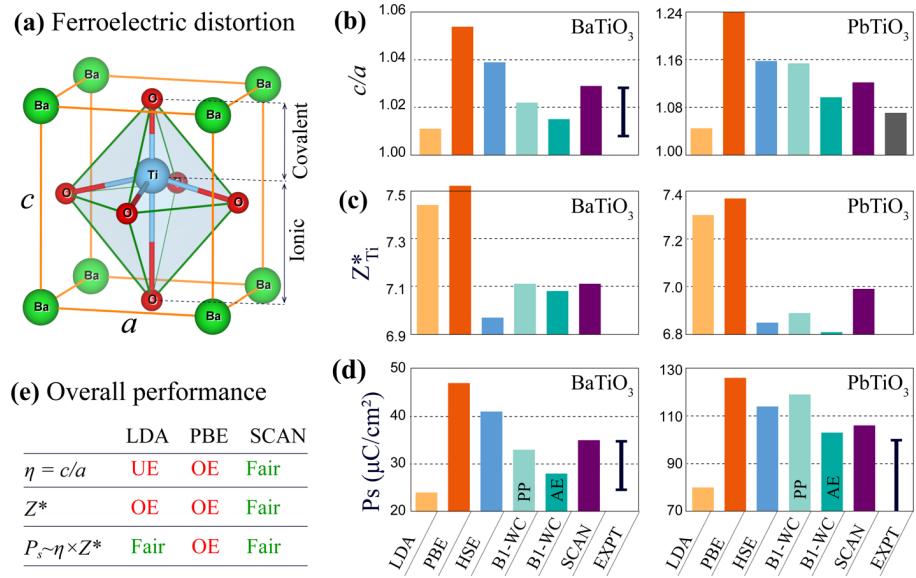

**Figure 9. Ferroelectric properties of $BaTiO_3$ and $PbTiO_3$.** (a) Polar structural distortion and the bonding interactions. (b) Lattice distortion, $\eta = c/a$. (c) Born effective charge, $Z^*$. (d) Ferroelectric polarization, $P_s$. The range of experimental values is indicated by error bars. PP and AE are pseudopotential and all-electron potential, respectively. (e) Performances of the LDA, PBE, and SCAN. UE and OE are short for *underestimation* and *overestimation*, respectively. Compiled from Ref. [31].

Ref. [31] revealed a fortuitous error cancellation in the LDA calculation. Firstly, the experimental measurements are subject to considerable uncertainties and can be significantly affected by external factors such as experimental temperature. Experimentally, $BaTiO_3$ exhibits a tetragonal phase between 278 and 353 K, transitioning to a cubic phase at higher temperatures. Consequently, the low-temperature value of $\eta = 1.026$ (for $BaTiO_3$) should be referenced for DFT calculations, and the reference $P_s$ should be around 35 and 100 μC/cm² for $BaTiO_3$ and $PbTiO_3$. Therefore, the LDA-predicted $\eta$ value is actually underestimated, and PBE's *super-tetragonality* issue may not be as critical as previously thought. Secondly, both LDA and PBE overestimate the Born effective charge ($Z^*$; Figure 9c) because these functionals tend to exaggerate the electronic polarizability due to the SIE [51,121]. According to a linear approximation, $P_s$ is represented by the product of the geometrical $\eta$ and electronic $Z^*$ (Figure 9e). Since LDA underestimates $\eta$ but overestimates $Z^*$, the resulting $P_s$ is surprisingly close to the experimental value due to the fortuitous error cancellation. In contrast, PBE's $P_s$ is overestimated due to the simultaneous overestimations of $\eta$ and $Z^*$. For all three properties, SCAN demonstrates the best alignment with the experimental data for the correct reasons.

In PbTiO$_3$, we explore the reduction of *super-tetragonality* error using the SCAN method compared to PBE. In the paraelectric phase, the electron localization function (ELF; see Figure 10a) shows a concentration of electrons around the oxygen anions, with Ti cations showing lesser electron density. This difference becomes evident when comparing SCAN and PBE calculations, where significant electron redistribution occurs ($\Delta n = n^{SCAN} - n^{PBE}$): Ti atoms lose electrons and oxygen atoms gain electrons, increasing the ionic interaction character in SCAN. Since ferroelectric distortion in Ti primarily arises from *pd* covalent bonding [122], the observed shift towards greater ionicity and reduced covalency directly reduces ferroelectric distortions, thus effectively mitigating the super-tetragonality error in SCAN. In the polarized phases, the stereochemically active Pb-$6s^2$ lone-pair electrons also play a crucial role (Figure 10b). The cap-shaped electron density leads to a non-centrosymmetric displacement of Pb atoms, enhancing the ferroelectric properties of PbTiO$_3$ compared to BaTiO$_3$ [119]. Notably, the calculated electron redistribution, $\Delta n$, around the Pb sites exhibits an interesting pattern: electrons migrate from the lower to the upper side of Pb. This electron movement counterbalances the cap-shaped electron distribution shown in the ELF and further helps reduce the super-tetragonality error in SCAN.

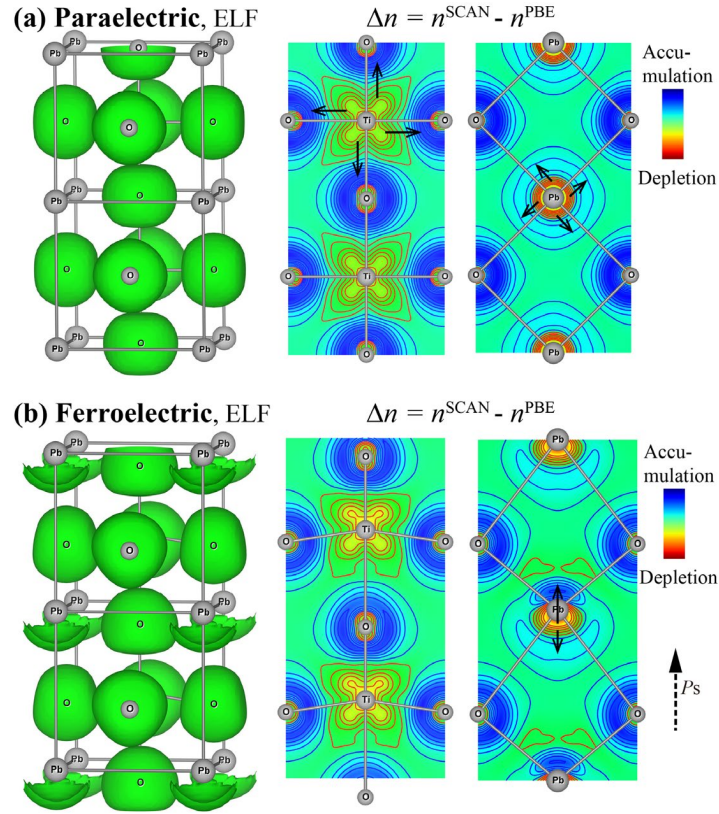

**Figure 10. Electron density analysis in PbTiO$_3$.** (a) The paraelectric phase. The left panel illustrates the electron localization function (ELF) with the green isosurfaces. The middle and right panels display the density changes on two (100) planes. Blue and red colors signify electron accumulation and depletion, respectively. The black arrows highlight the electron redistribution. (b) The ferroelectric phase. In the left panel, the cap-shaped ELF below the Pb ions highlights the Pb-$6s^2$ lone-pair electrons. The middle and right panels highlight the anisotropic electron distributions around Ti and Pb atoms. Cited from Ref. [123].

### 4.2.2. Multiferroics: Role of electron redistribution instead of significant *U* corrections

Multiferroic materials like BiFeO$_3$ [124,125] and YMnO$_3$ [126] present additional challenges for DFT calculations due to their open 3$d$ orbitals. In semiconductor simulations, establishing a positive bandgap is often a theoretical prerequisite. Both LDA and PBE significantly underestimate the bandgap of BiFeO$_3$ (Figure 11a), and a simple correction is to augment the functionals with a Hubbard-like *U* parameter. Interestingly, SCAN achieves results similar to PBE+*U* with *U* = 2 eV, including bandgap, ionic valence, local magnetic moments, and band structure (Figure 12a). This comparison suggests that SCAN inherently captures more "*U*-effects" than PBE, by an approximate margin of 2 eV [127]. For YMnO$_3$, which has an experimental bandgap of 1.35 eV [128], none of the considered functionals successfully open this bandgap (Figure 11b). The *U* values required vary significantly: PBE+*U* needs up to 5.0 eV, while SCAN+*U* requires a smaller value of 1.0 eV.

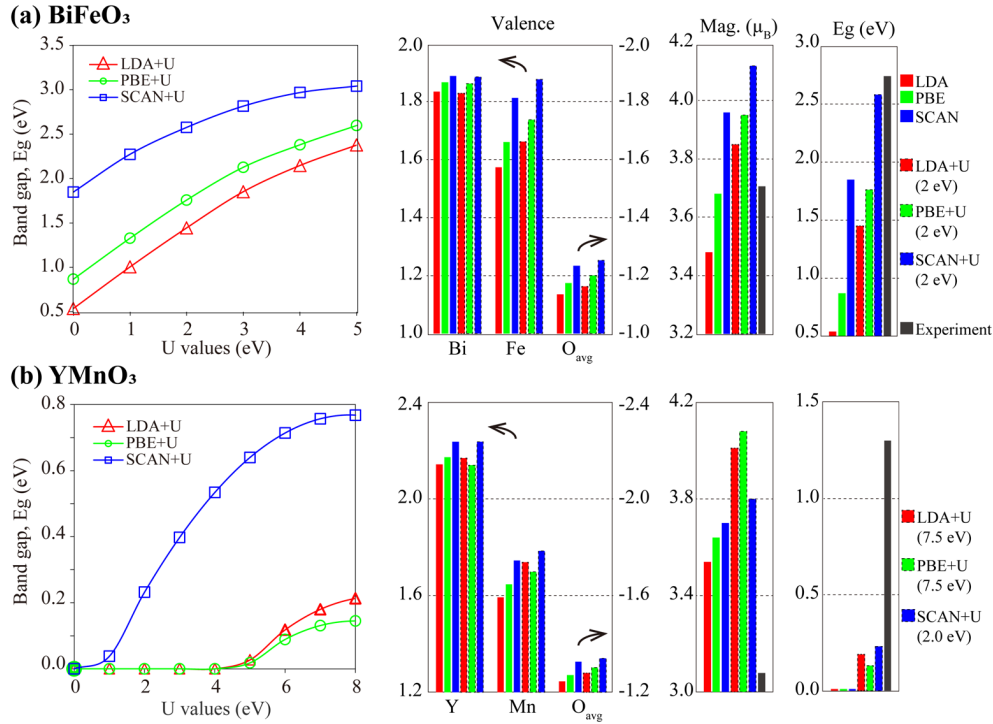

**Figure 11. Electronic properties of (a) BiFeO$_3$ and (b) YMnO$_3$.** The left panel displays the bandgaps with varying *U* values for the LDA+*U*, PBE+*U*, and SCAN+*U* methods. The second panel shows ionic valences; the oxygen valences are averaged if they exist in different chemical environments. The third and fourth panels detail the local magnetic moments and the bandgaps, respectively. Cited from Ref. [123].

Why does SCAN require a significantly smaller *U* parameter for YMnO$_3$? To investigate this puzzle, it is necessary to understand the unique electronic properties of YMnO$_3$ compared to BiFeO$_3$. As shown in Figure 12, the near-edge bands of YMnO$_3$ are highly dispersive, a distinctive feature absent in BiFeO$_3$. The top valence bands of YMnO$_3$ are derived from planar Mn-3$d$ orbitals (i.e., $d_{xy}$ and $d_{x^2-y^2}$) and O-2$p$ orbitals, exhibiting substantial σ-type overlap due to the directional orbitals (i.e., $pd$ hybridization). As discussed in previous sections, SCAN benefits from synergistic improvements in describing mixed states with both localized and delocalized electrons: it enhances $d$-orbital localization and effectively redistributes electrons between sites [123]. In YMnO$_3$, SCAN not only increases the localization of Mn-3$d$ orbitals, as seen from the increased local magnetic moment in Figure 11, but also captures more ionicity in the Mn-O bond, evidenced by the higher valency. This observation also applies to the bottom conduction bands, which derive from σ-type overlap between Mn-4$s$ and Mn-$d_{z^2}$ orbitals (i.e., $sd$ hybridization). Notably, SCAN better calculates the Mn-4$s$ orbitals, a feature unattainable by the +*U* method. As a result, SCAN simultaneously improves the simulation of the $s$, $p$, and $d$ states relevant to the bandgap, although a small correction of +*U* (~1.0 eV) is still needed to address the remaining delocalization error. The PBE+$U$ method faces a substantial limitation: it opens the bandgap exclusively by displacing the localized $d$-orbitals away from the band edge. Given the extensive $sd$ and $pd$ hybridizations, a $U$ value of 7.5 eV in PBE+$U$ is so substantial that it markedly distorts the band structure (Figure 12b) relative to experimental findings [123]. For instance, the Mn-3$d$ states are positioned so far from the band edge that they might not contribute effectively to the multiferroic properties.

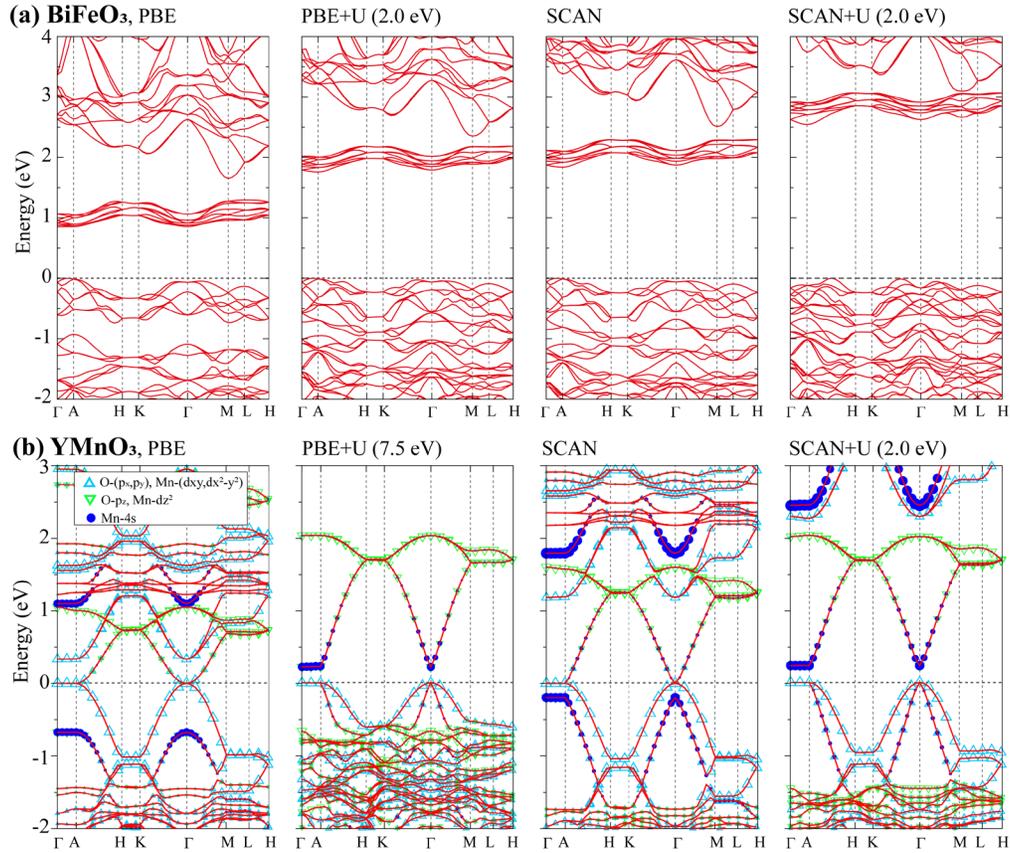

**Figure 12. Band structures of (a) BiFeO$_3$ and (b) YMnO$_3$.** For YMnO$_3$, wave functions are projected onto various orbitals. In the PBE+$U$ calculation, the conduction bands derived from the planar Mn $d$ orbitals (i.e., $dxy$ and $dx^2$-$y^2$) lie outside the plotting window due to an excessively high $U$ value of 7.5 eV. Cited from Ref. [123].

**4.2.3. Cuprates: Unified description of the pristine insulators and hole-doped metals**

Cuprates are highly complex ceramics that undergo an insulator-to-metal transition upon charge doping. The variability of Cu-3$d$ orbitals, which bridge the transition from a localized-immobile insulating state to an itinerant-conducting metallic state, presents significant difficulties for DFT simulations. While a widely cited failure for LDA is its inability to open a bandgap for the pristine insulating phase, hybrid functionals erroneously predict the charge-doped metal to remain an insulator [25]. Moreover, cuprates stand out from previous magnetic materials due to their small magnetic moments in the pristine phase, which becomes even weaker upon doping. For example, experimental measurements identified a magnetic moment of approximately 0.5 μB in La$_2$CuO$_4$ and YBa$_2$Cu$_3$O$_6$. However, LDA and PBE calculations underestimate the local magnetic moment in La$_2$CuO$_4$ [14] or fail to stabilize the magnetization in YBa$_2$Cu$_3$O$_6$ [129].

Figures 13a and 13b illustrate that SCAN simultaneously captures the insulating behaviors of La$_2$CuO$_4$ and the metallic characteristics of La$_{2-x}$Sr$_x$CuO$_4$ with $x$ = 0.25 [14,25,30]. We here analyze YBa$_2$Cu$_3$O$_{6+x}$ in greater detail as it is the first system to achieve a superconducting transition temperature exceeding the boiling point of nitrogen. Using SCAN, Figure 13c calculates the band structures with and without spin polarization for comparative analysis. The primary issue for the non-magnetic (NM) $YBa_2Cu_3O_6$ is the erroneous prediction of a metallic state. The metallic bands, derived from the planar Cu $dx^2-y^2$ orbitals and hybridized with O-2$p$ states, intersect the Fermi surface near points X, Y, and the Γ-S center of the Brillouin zone. Introducing spin polarization with G-type antiferromagnetism (AFM) markedly alters the band structure and induces a bandgap, energetically favoring this magnetic state by 115.9 meV per planar Cu atom. The AFM band structure is then unfolded back into the first Brillouin zone, clearly displaying bandgaps at the X, Y, and Γ-S center points.

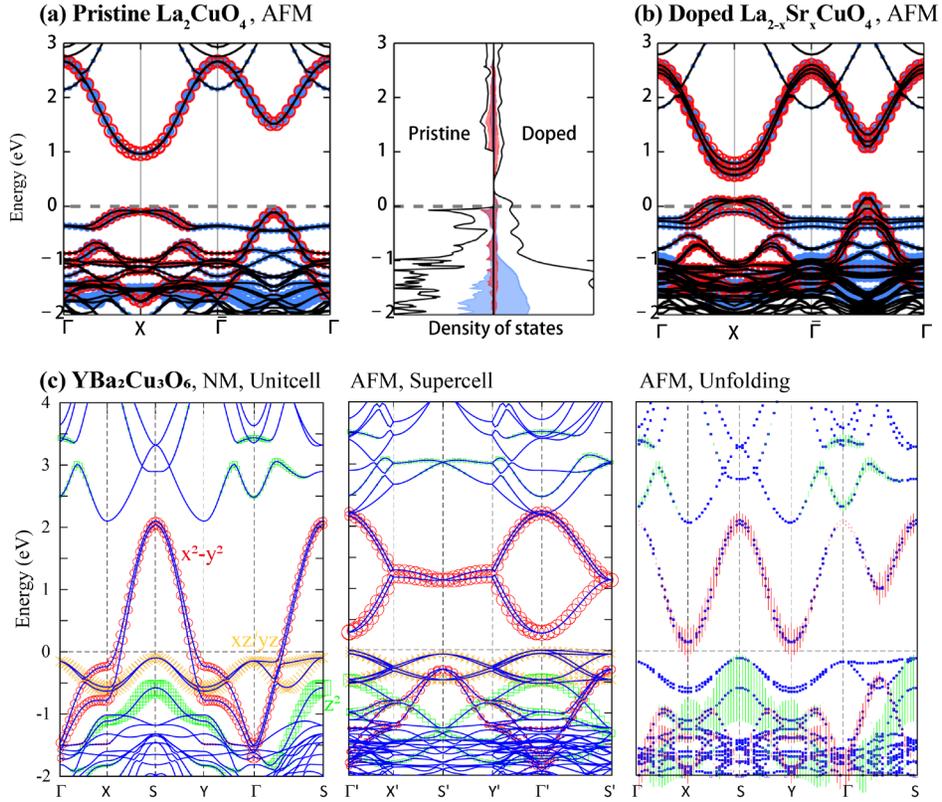

**Figure 13. SCAN-calculated band structure of cuprate oxides.** (a,b) Pristine $La_2CuO_4$ and Sr-doped $La_{2-x}Sr_xCuO_4$ with $x$ = 0.25. Oxygen $p_x+p_y$ orbitals are shown as blue dots, and copper $d_{x^2-y^2}$ as red circles. (c) Pristine $YBa_2Cu_3O_6$. The band structures are calculated for a non-magnetic (NM) state within a unitcell and a G-type antiferromagnetic (AFM) state within a 2×2×1 supercell. The contributions from planar Cu $d_{x^2-y^2}$ (red), chain Cu $d_{z^2}$ (green), and chain Cu $d_{xz/yz}$ (orange) states are overlaid on the band structures. Subplots (a) and (b) are cited from Ref. [14]; (c) is compiled from Ref. [129].

The ability of SCAN to accurately calculate the band structures highlights several strengths of this approach. First, SCAN's capacity to stabilize weak magnetization highlights its effectiveness in mitigating delocalization errors compared to PBE. Second, SCAN accurately captures the structural and orbital anisotropies in cuprates. Lastly, SCAN's adeptness at balancing these complexities is essential for accurately describing both the insulating and metallic phases.

### 4.2.4. Survey of perovskites from $d^1$ to $d^8$ electron configurations

Understanding and controlling the bandgaps of oxide perovskites, which constitute a significant group of functional materials, is crucial for their effective utilization. Due to partially filled 3$d$-shells, it is common practice to integrate DFT with an on-site $U$ to facilitate opening the bandgap. As the $U$ parameter originates from the Hubbard model, interpreting the gapping mechanism as a dynamic correlation has gained popularity. DFT might seem inadequate since it is a single-determinant mean-field-like approach that only incorporates static correlations. However, SCAN functional, without the $U$ parameter, has demonstrated systematic advancements in many perovskites [130].

Figure 14 surveys SCAN-predicted attributes of many perovskites, including their geometry, bandgap, magnetic moment, and phase stability properties [130]. Firstly, SCAN successfully opens the bandgap across all selected systems, from electron configurations from $d^1$ to $d^8$, in both antiferromagnetic and paramagnetic phases. These results align with experiments and reproduce trends in previous DFT+$U$ and DMFT (dynamical mean-field theory) simulations. Secondly, the authors explored structural symmetry breaking, including octahedra rotations, Jahn-Teller modes, and bond disproportionation, which are essential for reducing energy and ultimately facilitating bandgap formation. The SCAN-calculated lattice types generally agree with experimental observations, except for paramagnetic $YVO_3$ and $LaVO_3$ and antiferromagnetic $CaFeO_3$ for some specific reasons [130]. Lastly, SCAN yields magnetic moments comparable to those documented in the literature.

|  | $d$ filling | Sym. | Mag. | $\Delta E_{NM}$ (mev/f u) | Electronic properties | | Structural distortions | | |
|---|---|---|---|---|---|---|---|---|---|
|  |  |  |  |  | $E_g$ (eV) | $M_{3d}$ ($\mu_B$) | $Q_2^+$ (Å) | $Q_2^-$ (Å) | $B_{oc}$ (Å) |
| YTiO$_3$ | $t_{2g}^1 e_g^0$ | Pbnm | FM | -242 | 0.08 | 0.92 (0.84$^a$) | 0.03 | - | - |
|  |  | Pbnm | PM | -218 | 0.33 (1.20$^b$) | 0.84 | 0.02 (0.00$^c$) | - | - |
| LaTiO$_3$ | $t_{2g}^1 e_g^0$ | Pbnm | AFMG | -75 | 0.05 | 0.67 (0.46$^d$) | 0.06 | - | - |
|  |  | Pbnm | PM | -84 | 0.14 (0.20$^b$) | 0.78 | 0.04 (0.04$^c$) | - | - |
| YVO$_3$ | $t_{2g}^2 e_g^0$ | Pbnm | AFMG | -1052 | 0.89 | 1.77 (1.72$^e$) | 0.19 (0.14$^e$) | - | - |
|  |  | P-1 | PM | -978 | 0.55 (1.60$^f$) | 1.82 | NA | NA | - |
| LaVO$_3$ | $t_{2g}^2 e_g^0$ | Pbnm | AFMC | -864 | 0.78 | 1.78 (1.30$^g$) | 0.00 (0.01$^h$) | 0.09 (0.08$^h$) | - |
|  |  | P-1 | PM | -833 | 0.36 (1.10$^i$) | 1.80 | NA | NA | - |
| CaMnO$_3$ | $t_{2g}^3 e_g^0$ | Pbnm | AFMG | -2049 | 0.79 | 2.62 (2.64$^j$) | 0.01 | - | - |
|  |  | Pbnm | PM | -2036 | 1.46 (ins.) | 2.62 | 0.01 (0.04$^k$) | - | - |
| LaMnO$_3$ | $t_{2g}^3 e_g^1$ | Pbnm | AFMA | -1783 | 0.52 | 3.65 (3.70$^l$) | 0.28 | - | - |
|  |  | Pbnm | PM | -1758 | 0.30 (0.24$^m$-1.70$^n$) | 3.60 | 0.32 (0.30$^o$) | - | - |
| CaFeO$_3$ | $t_{2g}^3 e_g^0 + t_{2g}^3 e_g^2$ | P2$_1$ | AFMS | -1474 | 0.20 | 2.72-3.67 (2.48-3.48$^p$) | 0.00 | - | 0.13 |
|  |  | P2$_1$/n | PM | -1449 | 0.07 (0.25$^q$) | 2.66-3.72 | 0.00 (0.04$^p$) | - | 0.14 (0.18$^p$) |
| LaFeO$_3$ | $t_{2g}^3 e_g^2$ | Pbnm | AFMG | -1073 | 1.67 (2.10$^r$) | 3.94 (4.60$^r$) | 0.01 | - | - |
|  |  | Pbnm | PM | -936 | 0.52 | 4.04 | 0.01 (0.00$^s$) | - | - |
| YCoO$_3$ | $t_{2g}^6 e_g^0$ | Pbnm | NM | - | 1.48 (ins.$^t$) | - | 0.06 (0.05$^t$) | - | - |
| YNiO$_3$ | $t_{2g}^6 e_g^2 + t_{2g}^6 e_g^0$ | P2$_1$/n | AFMS | -353 | 0.92 | 1.41-0.00 (1.70-0.40$^u$) | 0.05 | - | 0.18 |
|  |  | P2$_1$/n | PM | -350 | 0.59 (0.20$^j$-1.00$^v$) | 1.38-0.28 | 0.05 (0.05$^w$) | - | 0.17 (0.13$^w$) |
| LaCuO$_3$ | $t_{2g}^6 e_g^2$ | R-3c | AFMG | -143 | 0.48 (?) | 0.70 (?) | - | - | - |

**Figure 14. Properties of oxide perovskites calculated using SCAN.** At low temperatures, spins are ordered into various long-range-ordered magnetic structures, including ferromagnetic (FM) and antiferromagnetic (AFM) with G, C, A, and S types. The high-temperature paramagnetic (PM) phase is modeled using a special-quasirandom-structure (SQS) to emulate magnetic disordering. NM denotes non-spin polarization. $\Delta E_{NM}$ (meV/f.u.) represents the energy difference between the spin-polarized and non-spin polarization solutions. $E_g$ denotes the bandgap, and $M_{3d}$ represents the local magnetic moment. Octahedral deformations are indicated by $Q_2^-$, $Q_2^+$, and $B_{oc}$ orthogonal modes. Experimental values are in parentheses, and *Ins.* signifies insulating phases. Cited from [130].

### 4.2.5. Section summary

This section illustrates how SCAN improves upon PBE for several types of complex materials, including ferroelectric, multiferroic, and cuprate materials. For ferroelectric $BaTiO_3$ and $PbTiO_3$, SCAN reduces the super-tetragonality error compared to PBE. In the multiferroic $YMnO_3$, SCAN requires a significantly smaller $U$ parameter to open the bandgap without distorting the band structure. For cuprate superconductors, SCAN accurately describes both the pristine insulating phase and the charge-doped metallic phase without relying on the empirical $U$. Additionally, we include a survey of oxide perovskites with electron configurations from $d^1$ to $d^8$, where SCAN surpasses PBE in terms of geometry, bandgap, magnetic moment, and phase stability. These studies demonstrate SCAN's capability to capture essential $d$-orbital behaviors in complex functional materials.

## 4.3. Complex electronic phases with nanoscale inhomogeneity

Magnetic materials can exhibit highly inhomogeneous electronic phenomena that extend to the nanoscale, potentially encompassing hundreds to thousands of atoms. Considering the computational effort involved, only semilocal density functional is a practical first-principle approach. Furthermore, the spatial inhomogeneities in these materials are frequently associated with fluctuations in magnetization and bonding interactions, which requires a functional capable of accurately capturing these subtleties. Previous studies have demonstrated that the SCAN (or r²SCAN) is particularly effective in elucidating the essential physics of these complex systems [15,129,131,132], and three examples are presented in this section.

### 4.3.1. Paramagnetic phase of monoxides with disordered magnetic moments

In Section 4.1.2, we adopted the G-type antiferromagnetic configuration within unitcell models when calculating the bandgaps of transition-metal monoxides. This approach assumes long-range magnetic order, typically established at low temperatures. However, the fundamental characteristic of Mott insulators is that their bandgap originates from electron-electron repulsions, independent of long-range magnetic ordering. Notably, the bandgap in a Mott insulator should persist even in the high-temperature paramagnetic phase, where local magnetic moments remain but are randomly oriented.

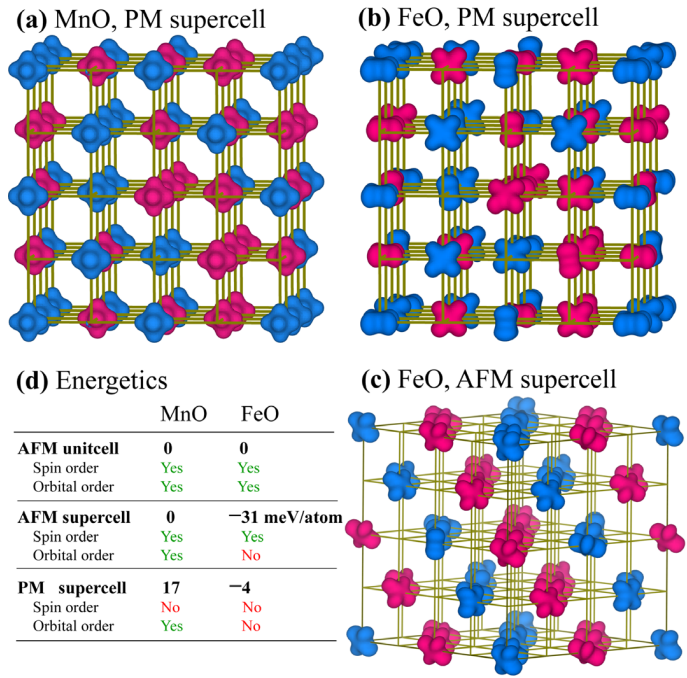

**Figure 15. Spin and orbital states in paramagnetic MnO and FeO simulated using SCAN.** (a) The paramagnetic phase of MnO. The blue and red isosurfaces represent the spin-up and spin-down densities, which are projected onto the valence-top $e_g^a$ orbitals (refer to Figure 4). Oxygen atoms are omitted for clarity. (b) The paramagnetic phase of FeO. The magnetic density features the shape of $a_{1g}$ orbital, distributed in a disordered pattern. (c) The antiferromagnetic phase of FeO in a supercell model. (d) Comparison of the electronic phases. The structural models contain 64 atoms. Compiled from Ref. [15].

Computationally, the widely-used non-magnetic model, which permits the coexistence of electrons with opposite spins on the same atomic site, often leads to an energetically unfavorable phase that does not accurately represent the actual magnetic states. To more accurately simulate the magnetic behavior, researchers have increasingly turned to models like the disordered-local-moment (DLM) [133-137] and special-quasirandom-structure (SQS) [15,130,138]. These models are designed to capture magnetic fluctuations in directions and magnitudes while maintaining zero net magnetization. Moreover, the utilization of paramagnetic supercell models raises important considerations regarding both computational efficiency and the accuracy of depicting magnetic atoms, which may become locally frustrated due to magnetic disordering. The SCAN meta-GGA functional has proven effective in this area: our results confirm that bandgaps are maintained across all four transition-metal oxides (i.e., MnO, NiO, FeO, CoO) in these supercell models [15].

Here, we select MnO and FeO as illustrative examples to demonstrate how the SCAN functional effectively opens bandgaps and also identifies a new ground state for FeO. For MnO, the disordering of magnetic moments is apparent from the distribution of magnetic densities shown in Figure 15a. The bandgap persists, although reduced to 0.80 eV for the paramagnetic phase compared to 1.67 eV for the antiferromagnetic phase [15]. This magnetic disordered phase is energetically less favorable than the antiferromagnetically ordered phase by 17 meV/atom (Figure 15d), primarily because the disordered magnetic moments capture less magnetic exchange energy. For FeO in the paramagnetic phase, the SCAN functional successfully opens the bandgap to 0.21 eV [15]. Additionally, the $a_{1g}$ orbitals spontaneously arrange into a randomly oriented pattern (Figure 15b), leading to a breaking of orbital symmetry that lower than the lattice symmetry. This *orbital-symmetry-breaking* is similarly observed in the antiferromagnetic phase when employing a supercell model (see Figure 15c), which yields a lower energy state by 31 meV/atom compared to the standard antiferromagnetic unitcell. This effect, highlighting the capability of the SCAN functional to detect orbital anisotropy, is unattainable with PBE because PBE inherently struggles to open a bandgap in FeO (refer also to Figure 8 and the related discussions).

### 4.3.2. Electronic stripe phase of cuprate and nickelate

While Section 4.2.3 is dedicated to the antiferromagnetic phase of cuprates, experimental evidence shows that doped cuprates tend to develop an electronic stripe phase extending to the nanoscale, which disrupts the translational symmetry of the underlying crystal lattice. A stripe phase represents a unique interplay of spin and charge order, characterized by domains of approximately (π,π) antiferromagnetic order separated by antiphase boundaries (APBs). These boundaries feature diminished magnetic moments and heightened charge densities [139]. Experimentally, the stable stripe phase consists of a $4a_0$ structural unit, where $a_0$ is the length of a $CuO_2$ square, with the APB centrally located on the Cu-Cu bond center [140,141].

SCAN accurately captures the fundamental characteristics of the $YBa_2Cu_3O_7$ stripe [129], as vividly illustrated in Figure 16a. The model features eight $CuO_2$ units aligned along the lattice-*a* direction, with two segments of electronic stripes, each denoted as $4a_0$ in the diagram. The presence of two stripes results from the periodic boundary conditions employed in the DFT simulation. Notably, the APBs are positioned at the center of the Cu-Cu bond, defining this as a bond-centered stripe phase. Furthermore, doped holes on the oxygen atoms (represented by the size of green spheres) tend to cluster around the APBs, while the local magnetic moments on Cu atoms (indicated by pink arrows) are more pronounced further from the APBs. The spatial separation between the doped holes and magnetic moments leads to atomic displacements from their original positions, triggering a Peierls-like breathing mode in the oxygen atoms (depicted by black arrows). These collective phenomena of hole-magnetization separation and atomic displacement could significantly alter the energetic landscape, as explored in Figure 16b. Without accounting for lattice contributions, the bond-centered stripe phase with $4a_0$ periodicity already displays an energetic minimum compared to other periodicities, aligning well with the experimentally determined stripe periodicity. Including lattice distortions further enhances the stability of the $4a_0$ stripe. The insights on lattice contributions provided by SCAN could complement the effective model Hamiltonian approach, which is popular for studying stripe physics but inadequately accounts for lattice effects.

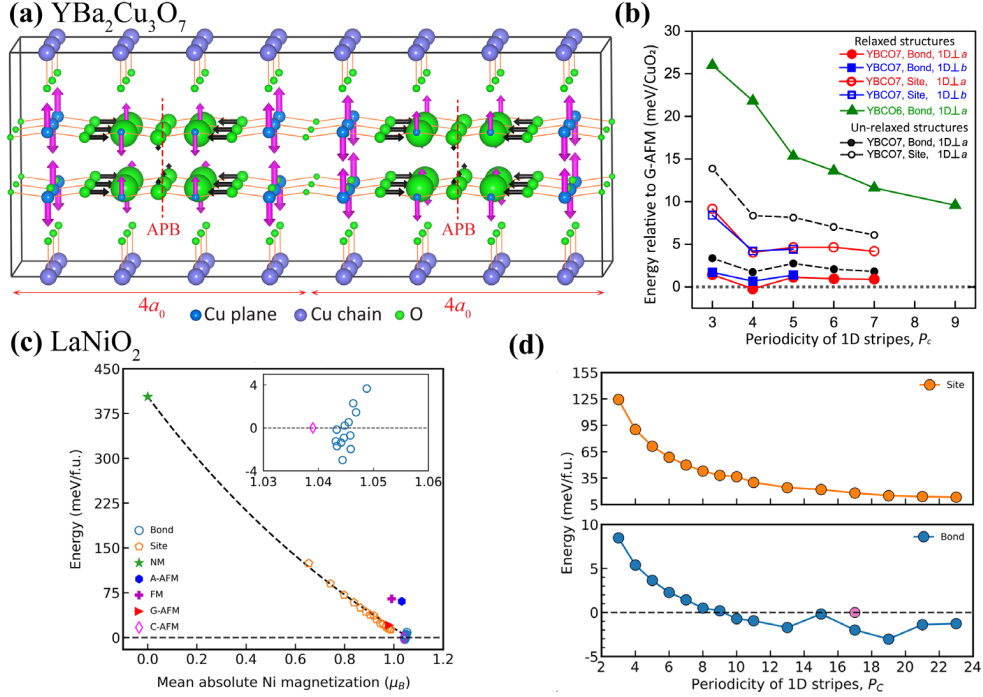

**Figure 16. Stripe phase of Cuprate and Nickelate oxides predicted by SCAN functional.** (a) Bond-centered stripe phase of $YBa_2Cu_3O_7$. The size of the green balls is proportional to the doped holes. Pink arrows denote the directions and sizes of the magnetic moments. Black arrows indicate the atomic displacements when forming the stripe phase. Y and Ba atoms are omitted for clarity. (b) Stripe energies relative to the G-type antiferromagnetic phase. (c) Energies of undoped infinite-layer $LaNiO_2$, plotted relative to the C-type antiferromagnetic phase as a function of the average magnetic moment of Ni atoms. (d) Relative energies plotted against their periodicity. The structural models contain up to ~400 atoms. Cited from [129,131].

The recent discovery of superconductivity in infinite-layer nickelates has reignited theoretical and experimental interest in the role of electronic correlations in their properties. Ref. [131] applied the SCAN functional to the undoped compound in the nickelate family, $LaNiO_2$. This study identified various competing low-energy stripe phases (Figure 16c), revealing that these stripe phases might possess lower energies than the long-ranged C-type antiferromagnetic phase. The local magnetic moments in these phases vary from approximately 0.6 to 1.0 $\mu_B$, all effectively captured by the SCAN functional. The near-degeneracy among the stripe phases suggests that the actual ground state could fluctuate between these configurations, aligning with experimental observations. Figure 16d further demonstrates that the bond-centered stripes are more stable than the site-centered ones, a characteristic consistent with observations in the cuprates. Notably, the bond-centered stripes increase in stability as their periodicity exceeds $10a_0$, a trend that diverges from that observed in the doped cuprates. The SCAN simulations predict that strong electronic correlations and electron-phonon coupling effects are crucial in $LaNiO_2$.

### 4.3.3. Kondo-like spin compensation phase in off-stoichiometric Heuslers

In efforts to enhance thermoelectric efficiency, the integration of magnetism remains comparatively underexplored and presents an ongoing intellectual challenge. Recently, a group of vacancy-filling Heusler alloys [132,142], such as $Nb_{0.75}Ti_{0.25}FeCr_xSb$ ($0 \leq x \leq 0.1$) with a partial and random filling of the 4c and 4d Wyckoff positions, has demonstrated promising capabilities to incorporate magnetism and improve thermoelectric performance. In the experiment, the system of $Nb_{0.75}Ti_{0.25}FeCr_{0.10}Sb$ achieves a remarkable thermoelectric figure-of-merit ($zT$) 1.21 at 973 K [132].

The formula of $Nb_{0.75}Ti_{0.25}FeCr_{0.10}Vac_{0.90}Sb$ ('Vac' stands for 'vacancy') inherently suggests a complex structure due to occupational disordering. Constructing a realistic theoretical model that fully captures the complexities of occupational disordering and magnetic phenomena might seem formidable. Nonetheless, the r²SCAN functional has identified intriguing Kondo-like magnetic features, agreeing well with the experimental results [132]. Here, we reproduce some essential theoretical results. Figure 17a shows a supercell model of Nb$_{75}$Ti$_{25}$Fe$_{100}$Cr$_{10}$Vac$_{90}$Sb$_{100}$, comprising 310 atoms and 90 vacancies, which adequately represents the occupation randomness. This structure includes 100 4b Wyckoff positions occupied by a random alloy of 75 Nb and 25 Ti atoms; There are 200 4c+4d Wyckoff positions available, but they are partially and randomly occupied by insufficient atoms (i.e., 100 Fe and 10 Cr atoms). In this supercell model, determining the ground-state spin configuration is also a significant challenge due to the extensive configuration space. However, r²SCAN simulations reveal a straightforward principle of magnetic coupling. As depicted in Figure 17b, Model-1 and Model-2 have the same crystal structures but start with different spin configurations. Structural relaxation and the magnetic states are then updated self-consistently. While the two models display markedly different magnetic moments, they have near-degenerate energies with a tiny difference of $\Delta E \sim 0.72$ meV/atom. This small $\Delta E$ is closely associated with Kondo-like spin compensation, as illustrated in Figure 17c: Cr sites with significant magnetization are tightly surrounded by opposite spin densities derived from less localized Fe and Ti orbitals. This magnetic behavior increases the effective mass of the electrons, which in turn enhances the Seebeck coefficient and consequently improves the thermoelectric performance.

In this study, several critical factors necessitate using an advanced DFT functional like r²SCAN. First, the structural model's size demands a functional that is computationally efficient. Second, the model features a mix of strong magnetization (e.g., $> 3\ \mu_B$ for Cr) and weaker magnetic elements (e.g., Fe, Ti, and Nb), with magnetic densities that are highly non-spherical around the atomic sites. These distinctive characteristics may challenge conventional methods such as PBE and PBE+$U$. While PBE struggles to stabilize the weak magnetization, the PBE+$U$ method faces difficulties in finding the optimal $U$ values to address the inhomogeneity.

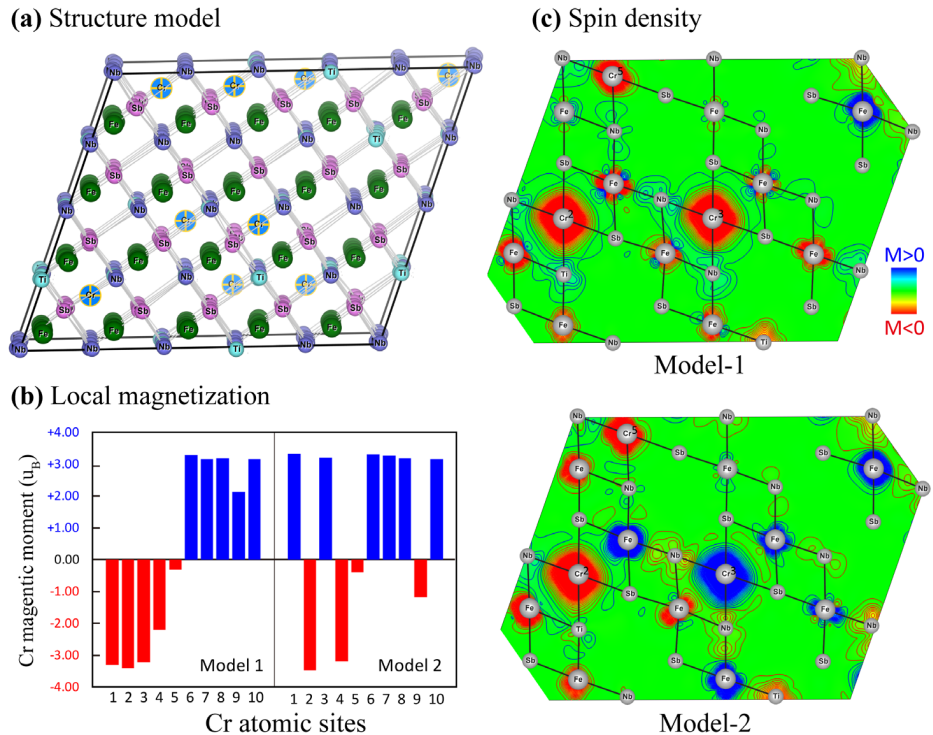

**Figure 17. Magnetic properties of off-stoichiometric Heusler Nb$_{75}$Ti$_{25}$Fe$_{100}$Cr$_{10}$Vac$_{90}$Sb$_{100}$ simulated by the r²SCAN functional.** (a) Supercell model showing occupation disorder. (b) Local magnetization at Cr sites. (c) Kondo-like spin compensation. The two models have different spin configurations but near-degenerate energies. The model has 310 atoms. Compiled from Ref. [132].

### 4.3.4. Section summary

This section examines three materials characterized by nanoscale inhomogeneity and fluctuating magnetic moments (directions and magnitudes) across sites. In FeO, SCAN calculations have sustained a bandgap in a spin-disordered paramagnetic state, effectively capturing the essence of Mott insulation. Additionally, these calculations have revealed a new ground state that showcases long-range magnetic order alongside disordered orbitals. In high-temperature superconductors, SCAN calculations successfully replicate the well-documented stripe phase in doped cuprates and intriguingly predict a similar stripe phase in the undoped nickelate material $LaNiO_2$. We also examine the thermoelectric Heusler $Nb_{0.75}Ti_{0.25}FeCr_xSb$, where intrinsic magnetic properties substantially bolster thermoelectric performance. The r$^2$SCAN simulations efficiently and accurately capture the Kondo-like spin compensation between strongly and weakly magnetized elements. In sum, the three studies demonstrate the robustness of SCAN and r$^2$SCAN in characterizing inhomogeneous electronic structures, where conventional methods like PBE and PBE+$U$ may encounter difficulties.

## 5. Energetics and lattice dynamics

In condensed matter physics and materials science, the energetic stability and lattice vibrational properties are fundamental to understanding the materials. Given the enhancements that SCAN has brought to electronic property predictions, it is logical to expect that SCAN will also yield more accurate forecasts of energetics and the associated lattice dynamics. Many studies have evaluated SCAN's performance concerning TMCs [16,17,19,20,22,27,29,105], and we provide a concise overview of some assessments.

### 5.1. High throughput calculation of material formation enthalpy

Evaluating chemical stability, i.e., whether a stoichiometry will persist in some chemical environment, and structure selection, i.e., what crystal structure a stoichiometry will adopt, is critical to predicting materials synthesis, reactivity, and physical properties. The *chemical accuracy* of formation enthalpies ($\Delta H_f$) is typically quoted as 1 kcal/mol (0.04 eV/atom), achievable in experiments. Due to the diversity of chemical degrees of freedom, it is a significant challenge for DFT to address both facets of the stability problem within a reasonable computational budget. While PBE is the dominant approach, the errors in formation enthalpy predictions by PBE are typically around ~0.2 eV/atom. There are three primary sources of PBE's error: SIE, incomplete error cancellation between the target compound and the elemental references, and the lack of van der Waals (vdW) interactions.

Reference [29] demonstrated that for 102 main-group compounds, the formation enthalpy calculated using the SCAN method has a mean absolute error (MAE) of 0.084 eV/atom, which is approximately 2.5 times more accurate than the MAE of 0.21 eV/atom observed with the PBE method. Additionally, SCAN's MAE for 21 main-group oxides is 0.038 eV/atom, notably achieving the *chemical accuracy*. Regarding predicting the correct ground-state structures, SCAN substantially outperforms PBE by decreasing the incidence of structural prediction errors from 12% to only 3% with a tolerance of 0.01 eV/atom.

We now focus on TMCs, in which SIE dramatically reduces the efficacy of semi-local density functionals. Specifically, for the formation enthalpies of 98 transition-metal binaries, Figure 18a shows that the MAE for SCAN is 0.122 eV/atom. This error is significantly larger than that found for main-group compounds, highlighting challenges specific to transition metals. Despite this, SCAN still presents an improvement over PBE, reducing the error by about 0.08 eV/atom, which represents approximately 40% of the total error attributed to PBE. Furthermore, in an analysis involving 106 transition-metal binaries across 1336 structures, SCAN markedly improves the accuracy of structural selection compared to PBE, as illustrated in Figure 18b.

Recent research [19] has conducted more comprehensive investigations into 1015 solid materials. This study also utilized r$^2$SCAN and included dispersion corrections. Figure 18c shows the main results for various database subsets, including the TMCs. Across the entire dataset, r$^2$SCAN achieves the lowest MAE at 92 meV/atom among all tested density functionals, while SCAN exhibits a slightly higher MAE of 107 meV/atom. For intermetallics, PBE proves to be the most accurate, predicting formation enthalpies with an MAE of 70 meV/atom; however, SCAN and SCAN+rVV10 show significantly higher errors, recording 148 meV/atom and 150 meV/atom, respectively. The inferior performance of SCAN in this context was initially reported in Ref. [17] and will be further discussed in Section 6.3.3. In the case of TMCs, r$^2$SCAN leads with the lowest MAE of 97 meV/atom, followed by r$^2$SCAN+rVV10 at 108 meV/atom, SCAN at 114 meV/atom, SCAN+rVV10 at 122 meV/atom, and PBE at 137 meV/atom.

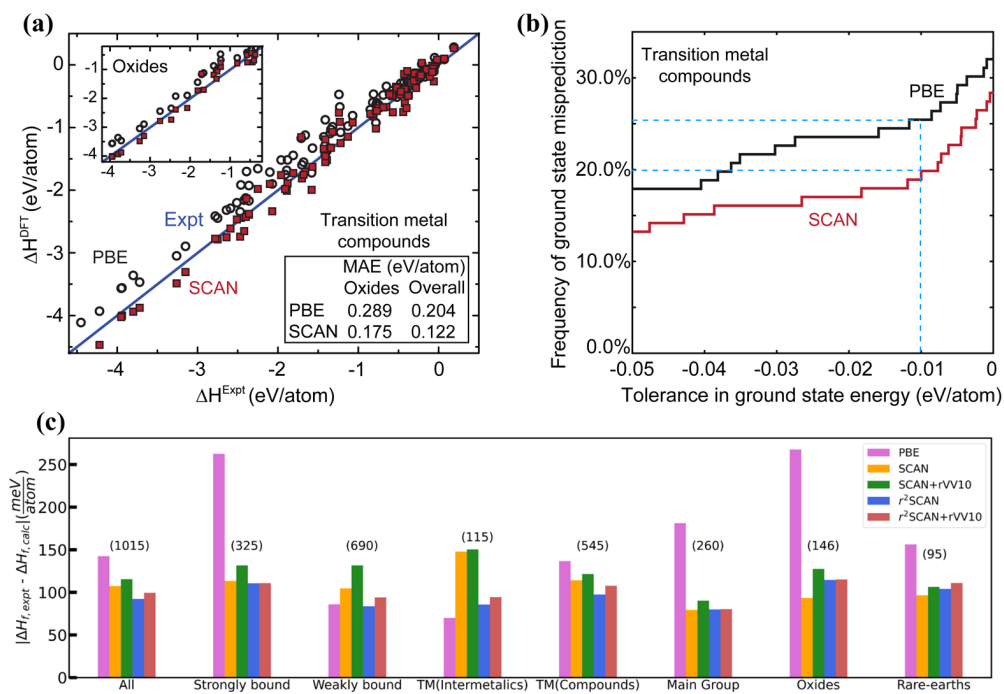

**Figure 18.** Formation energy and structure selection of transition-metal compounds. (a) The formation enthalpy of 98 binaries, and oxides in particular. (b) The probability of incorrect structure selections across 106 compounds for a range of tolerances. (c) Mean absolute errors of the calculated formation energy. "All" refers to the entire set of 1015 solids; "strongly bound" solids are those with experimental formation enthalpies ranging from –4 to –1 eV/atom; "weakly bound" solids are characterized by $|\Delta H_{\text{expt}}| < 1$ eV/atom. Transition-metal systems are categorized into TM(Intermetalics), which consist solely of transition metals; TM(Compounds) includes other elements. Parenthetical numbers above each set of bars indicate the number of compounds in each subset. Cited from [19,29].

## 5.2. Discriminating energetically near-degenerate polymorphs of TiO$_2$

The challenge of structural selection originates from distinguishing electronically similar phases that compete for the more stable ground state. Additionally, since these competing phases differ structurally but not chemically, their formation enthalpies often vary on a very fine energy scale. To enhance our understanding of the improvements and persistent challenges associated with the SCAN functional, we have chosen TiO$_2$ for a detailed examination [143].

TiO$_2$ exists in six polymorphs: rutile, anatase, brookite, β-TiO$_2$, α-PbO$_2$, and baddeleyite. Among these, rutile and anatase are the leading contenders for the ground state. While nanostructured TiO$_2$ predominantly exhibits the anatase phase, the consensus among most experimental studies is that rutile is thermodynamically more stable for the bulk crystalline TiO$_2$. Accurately determining the relative stability between anatase and rutile poses significant challenges for DFT functionals. Figure 19c presents the calculated energetic orderings of TiO$_2$ polymorphs alongside experimental data. The PBE functional incorrectly predicts anatase to be significantly more stable than it actually is, with a calculated energy difference $\Delta E = E_{\text{anatase}} - E_{\text{rutile}} = -95$ meV/formula. Although SCAN functional improves upon this prediction, it still does not align accurately with experimental observations, indicating anatase to be more stable than rutile by $\Delta E = -26$ meV/formula. Given the relatively localized nature of Ti-3$d$ orbitals, augmenting the DFT functionals with +$U$ corrections is common practice to better handle the SIE. To correct the energetic ordering between anatase and rutile, a significantly high $U$ value of 6 eV is required in the PBE+$U$ approach, while it is reduced to 1.5 eV in SCAN+$U$.

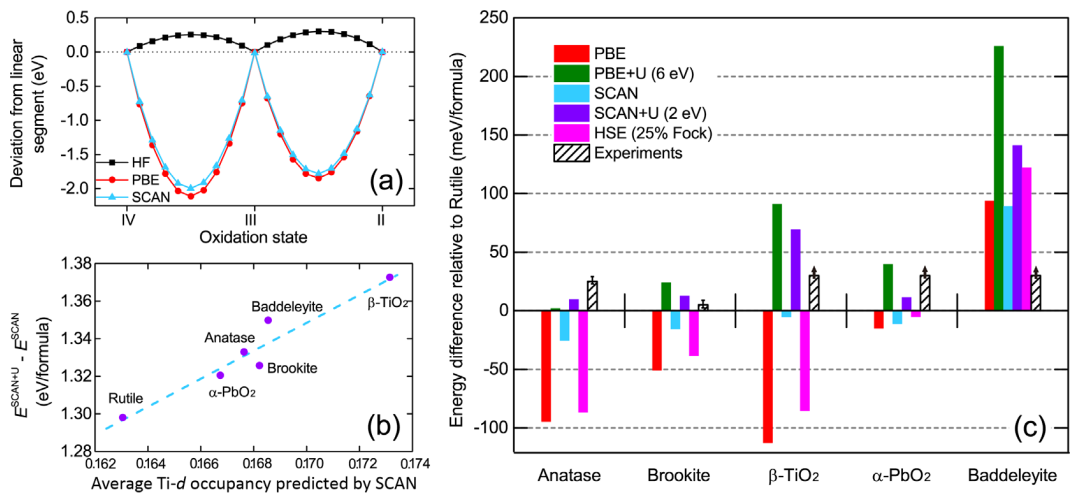

**Figure 19. Calculation of the relative stabilities of six TiO$_2$ polymorphs.** (a) Deviation from the linear energy segment as a function of the occupation number of the isolated Ti ions. (b) Energetic correction by the Hubbard-like $U$ (i.e., $E^{SCAN+U} - E^{SCAN}$, $U = 2$ eV) with respect to the average Ti-3$d$ sub-orbital occupancies (the unit is electron per sub-$d$-orbital per spin-channel), which are calculated by averaging the total 3$d$ occupancy over the five Ti-3$d$ sub-orbitals with both spin directions. (c) Relative stability from various theoretical approaches. Cited from [143].

How does the corrective $U$ help the functionals correctly capture the stability of TiO$_2$, and why is it significantly smaller for SCAN than for PBE? Figure 19a (similar to Figure 6b) illustrates the energy deviations of isolated Ti ions from the linear trends defined by formal integer valences (i.e., +2, +3, and +4 oxidation states) in relation to fractional electron occupations [39,144]. The largest deviation occurs when the state is half an electron away from these integer valences—a scenario that presents a significant challenge for all semi-local density functionals. Indeed, Figure 19b demonstrates that the 3$d$ orbitals of TiO$_2$ across all phases are fractionally occupied, with rutile and β-TiO$_2$ showing the lowest and highest occupations, respectively. The corrective $U$ adjustment decreases the fractional occupancy by approximately 0.015 electrons for each sub-$d$-orbital of each spin, effectively localizing the Ti-3$d$ electrons. This correction is the weakest for rutile and the strongest for β-TiO$_2$, resulting in an almost linear relationship with the Ti-3$d$ occupations. Revisiting Figure 19a, the less pronounced downward deviation in SCAN compared to PBE indicates weaker SIE in SCAN, necessitating a smaller $U$ correction.

## 5.3. Lattice vibrations of magnetic materials: Fe, NiO, and YBa$_2$Cu$_3$O$_6$

Lattice vibrations or phonon frequencies, derived from the second derivative of the potential energy with respect to atomic positions, are crucial for understanding a broad spectrum of material properties. For instance, the role of lattice dynamics in cuprate superconductors remains a hotly debated topic. DFT remains the most accessible first-principles method, yet it requires highly accurate calculations of forces and force constants to correctly evaluate dynamical properties. Such precision requires reliable and efficient DFT functionals, especially for transition-metal materials where accurately calculating electronic properties is already challenging. Prior evaluations revealed that the original SCAN functional struggled with numerical issues in complex materials calculations, sometimes incorrectly predicting dynamical instability in phonon simulations [22]. These issues have been addressed in the r$^2$SCAN functional [11], which enhances numerical stability while preserving the accuracy of SCAN, as discussed in section 6.3.1.

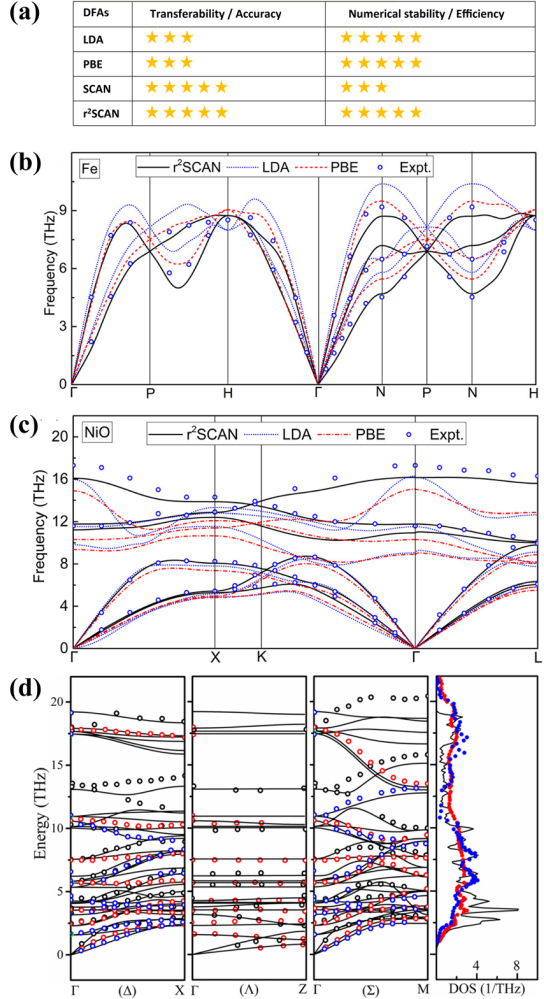

**Figure 20. Lattice vibrational properties of transition-metal materials.** (a) Performance summary of four functionals. (b) Phonon dispersion for ferromagnetic Fe. (c) Phonon dispersions for antiferromagnetic NiO. (d) Phonon dispersions and density of states for antiferromagnetic $YBa_2Cu_3O_6$. Black lines represent the $r^2$SCAN results, while the circles and dots indicate experimental data. Cited from [22,27].

The overall performance of four popular DFAs in phonon simulations is summarized in Figure 20a: $r^2$SCAN outperforms the other methods in accuracy and efficiency. We select four transition-metal materials of ferromagnetic Fe [22], antiferromagnetic NiO [22], and $YBa_2Cu_3O_6$ [27], representing comprehensive scenarios where assessing the reliability and effectiveness of phonon calculations is critical. Figure 20b compares calculated and experimental phonon dispersions for elemental Fe. The differences between PBE and $r^2$SCAN are subtle, with each showing areas where they align closely with experimental data. The N high-symmetry point is notably challenging for all functionals. The $r^2$SCAN functional, however, shows superior performance for the lowest band along the Γ−N−P−N−H path. PBE tends to be more precise at higher frequencies, aligning better with the experimental results for the middle and higher bands. We mention that despite $r^2$SCAN's tendency to overestimate the magnetic moment of Fe [24], this does not significantly impact the accuracy of its phonon dispersion predictions. Figure 20c focuses on the phonon dispersion for NiO, where $r^2$SCAN demonstrates significant improvements, particularly in accurately capturing the high-frequency optical bands—a task at which both LDA and PBE falter. These enhancements are likely attributed to the SIE reduction in $r^2$SCAN, an important factor considering the involvement of Ni-3$d$ shells. In Figure 20d, the phonon properties of the cuprate $YBa_2Cu_3O_6$ are explored. The $r^2$SCAN simulations closely match the experimental dispersion curves of critical phonon modes. The analysis also uncovers notable magnetoelastic coupling in the high-energy Cu-O bond stretching optical branches. This coupling shows an enhancement in magnetization related to the softer nonmagnetic phonon bands, highlighting the interplay between magnetic and lattice dynamics in this material.

### 5.4. Section summary

This section evaluates the performance of SCAN and r$^2$SCAN in calculating properties that critically depend on precise energy calculations: the formation enthalpy of compounds, the relative stability of a compound's polymorphs, and lattice vibrations from phonon dispersion. High-throughput simulations of formation enthalpies demonstrate that SCAN and r$^2$SCAN significantly mitigate the errors associated with the PBE functional. Furthermore, SCAN and r$^2$SCAN have shown superior ability in assessing the relative stability of polymorphs, a process that involves distinguishing subtle energetic differences between different structures but with identical compositions. For example, in predicting the energy hierarchy of six $TiO_2$ polymorphs, SCAN markedly outperforms PBE, although it still requires additional adjustments to resolve the remaining SIE. Finally, r$^2$SCAN, known for enhancing numerical stability while maintaining the accuracy of SCAN, has proven both reliable and efficient in calculating phonon frequencies, as demonstrated in magnetic materials of Fe, NiO, and $YBa_2Cu_3O_6$.

## 6. Other issues and discussions

### 6.1. Materials with rare-earth *f*-electrons

Materials containing rare-earth elements can exhibit fascinating electronic and magnetic phenomena, primarily due to the extremely localized nature of the *f*-orbitals. Heavy-fermion compounds are a notable example, where the strong interactions between the localized *f*-electrons and the itinerant *s*-electrons lead to the Kondo effect. Accurately describing the localization-delocalization interplay remains a significant challenge for DFT functionals.

Reference [26] investigated the ground-state electronic structure of $SmB_6$, a prototypical heavy-fermion compound. Although this material is electrically insulating, it exhibits metallic characteristics such as quantum oscillations. The study identified several competing magnetic phases closely aligned in energy, indicating significant spin fluctuations. The computed band structure, crystal-field splitting, the heavy effective electron mass, and the large specific heat all agreed with experimental data. The superior performance of SCAN in this context is attributed to its enhanced ability to handle SIE.

However, another study on rare-earth oxides ($EuTiO_3$, $GdTiO_3$, $PrCrO_3$, and $DyFeO_3$) reported less satisfactory results [145]. While SCAN can occasionally predict the insulating character of some compounds accurately, it frequently fails to correctly predict the band-edge orbital character of insulators. This failure is ascribed to the functional's insufficient correction of SIE, leading to an underestimation of Hund's splitting associated with 4*f* states. As a result, unlike the pervasive improvements achieved in TMCs, the universal applicability of the SCAN functional for rare-earth materials remains elusive.

### 6.2. Long-range van der Waals interactions: SCAN+rVV10 and r$^2$SCAN+rVV10

Atomically thin two-dimensional materials usually exhibit significant long-range van der Waals (vdW) interactions. Although SCAN and r$^2$SCAN effectively capture short- to intermediate-range interactions, they require augmentation with specific treatments to adequately address long-range vdW interactions. SCAN+rVV10 represents a "best-of-both-worlds" approach, which effectively integrates the SCAN meta-GGA with the rVV10 nonlocal correlation functional [146]. As of its development, SCAN+rVV10 is the only vdW density functional that delivers excellent results for interlayer binding energies and spacings, as well as intralayer lattice constants across 28 layered materials. Its versatility for various kinds of bonding is further demonstrated by its good performance for 22 interactions between molecules, the cohesive energies and lattice constants of 50 solids, the adsorption energy and distance of a benzene molecule on coinage-metal surfaces, the binding energy curves for graphene on metal surfaces, and the rare-gas solids.

For the r$^2$SCAN functional, refitting the rVV10 functional results in the r$^2$SCAN+rVV10 vdW density functional [147]. Molecular tests demonstrate that r$^2$SCAN+rVV10 outperforms its predecessor, SCAN+rVV10, in both efficiency and accuracy. This good performance is also found in lattice-constant predictions. Compared with benchmark results from higher-level theories or experiments, r$^2$SCAN+rVV10 yields excellent interlayer binding energies and phonon dispersions in layered materials.

### 6.3. Remaining problems for SCAN and r$^2$SCAN

#### 6.3.1. Numerical issue of SCAN and its resolution by r$^2$SCAN

After the publication of the original SCAN functional [10], many studies reported numerical difficulties, especially in TMCs like FeO [15]. These problems arise from using the dimensionless orbital indicator, $\alpha$, which tunes the functional's performance for specific local chemical environments. While $\alpha$ allows SCAN to meet exact constraints that would be contradictory at the GGA level, it can also induce numerical sensitivity and divergences in the XC potential. The rSCAN functional [91] was designed to manage SCAN's numerical challenges while making minimal changes to the original SCAN. This adaptation significantly improves numerical stability and facilitates pseudopotential generation. Although initial testing indicated that rSCAN preserved both the accuracy and transferability of SCAN, further evaluations revealed a loss in transferability, particularly affecting the accuracy of atomization energies. The r$^2$SCAN functional maintains the numerical stability of rSCAN while restoring the transferable accuracy of the original SCAN [11]. These improvements have been demonstrated across materials with varying bonding characteristics. Notably, r$^2$SCAN shows excellent performance in phonon calculations, which depend on precise and efficient computations to avoid numerical instabilities or inaccuracies [22].

#### 6.3.2. Overestimation of Fe's magnetic moment

Despite the widespread improvements offered by SCAN over PBE, Fu and Singh [24] observed that SCAN fails to accurately describe the stability and properties of various iron phases, attributing this failure to an overestimated tendency toward magnetism and an exaggeration of magnetic energies. Table 2 compares the magnetic moments of four transition metals as calculated using various functionals. Both SCAN and r$^2$SCAN predict larger-than-expected magnetic moments for body-centered cubic (bcc) iron, hexagonal close-packed (hcp) cobalt, and face-centered cubic (fcc) nickel. Most notably, while PBEsol accurately reproduces the magnetic moment of iron, both SCAN and r$^2$SCAN overestimate it by approximately 30%.

Trickey et al. propose a method to eliminate explicit orbital dependence from SCAN and r$^2$SCAN by converting them into Laplacian-dependent functionals, named SCAN-L [148] and r$^2$SCAN-L [149], respectively. Interestingly, the issue of overmagnetization in Fe, Co, and Ni is resolved in both SCAN-L and r$^2$SCAN-L. Another deorbitalized variant of SCAN and r$^2$SCAN, known as OFR2 and developed by Kaplan and Perdew [150], also accurately predicts the magnetic moments. According to the authors [150], OFR2 is recommended for an orbital-free description of solids and liquids, particularly for *sp* or *sd* metals; however, SCAN and r$^2$SCAN continue to be recommended for achieving the best accuracy in molecules and nonmetallic condensed matter.

**Table 2. Local magnetic moment of transition metals.** Saturation magnetizations ($\mu_B$/atom) of four elemental solids at their respective experimental lattice constants. Reference a [150] and b [149].

|  | bcc Fe | hcp Co | fcc Ni | bcc V |
| --- | --- | --- | --- | --- |
| PBEsol [a] | 2.094 | 1.570 | 0.620 | / |
| SCAN [b] | 2.60 | 1.80 | 0.78 | 0.57 |
| SCAN-L [b] | 2.05 | 1.63 | 0.67 | 0.00 |
| r$^2$SCAN [b] | 2.63 | 1.77 | 0.74 | 0.03 |
| r$^2$SCAN-L [b] | 2.27 | 1.67 | 0.69 | 0.06 |
| OFR2 [a] | 2.12 | 1.63 | 0.66 | / |
| Experiment [a] | 1.98-2.13 | 1.52-1.58 | 0.52-0.57 | / |

### 6.3.3. Formation energies of intermetallic compounds

As discussed in Section 5.1, SCAN underperforms compared to PBE in calculating the formation energy of intermetallic compounds, a finding initially reported by Isaacs and Wolverton [17]. To understand SCAN's inferior performance, Ref. [17] categorized compounds into strongly and weakly bound groups. They then analyzed the exchange behavior in these two categories. A distinct exchange behavior was observed in the weak bonding regime, contributing to SCAN's underperformance in calculating properties of weakly bound compounds, such as intermetallic materials.

### 6.3.4. Geometry of chalcopyrite CuInSe$_2$

Recent research [108] has demonstrated that semi-local functionals, including SCAN and r$^2$SCAN, face challenges in accurately predicting the geometric structure of CuInSe$_2$, a well-known photovoltaic material. CuInSe$_2$ is characterized by a diamond-like structure comprising Se-Cu$_2$-In$_2$ tetrahedrons (Figure 21a). The Se atoms are displaced from the centers of these tetrahedrons due to differences between Cu-Se and In-Se bond lengths, and the displacement is quantified by the $\mu$ parameter (defined in Figure 21d). Experimental measurements of this parameter show significant variability, with theoretical predictions generally falling below the lower experimental bounds. This $\mu$ parameter is crucial as it significantly influences the material's electronic properties: an experimental geometry with $\mu = 0.230$ results in a semiconducting band structure (Figure 21b), whereas the geometry optimized with r$^2$SCAN, which has $\mu = 0.219$, incorrectly predicts a semi-metallic structure (Figure 21c). Previous studies have indicated that Cu-3$d$ orbitals can exhibit strong delocalization errors with commonly used DFT functionals [151]. Although introducing a $+U$ correction to r$^2$SCAN was intended to address this delocalization error, it has only resulted in limited improvements. In contrast, the HSE06 hybrid functional has demonstrated reliability in calculating all geometric parameters, including the anion displacement (Figure 21d). Further investigation is necessary to fully understand the computational challenges associated with CuInSe$_2$ and related materials such as Cu$_2$S (refer to Section 4.1.1).

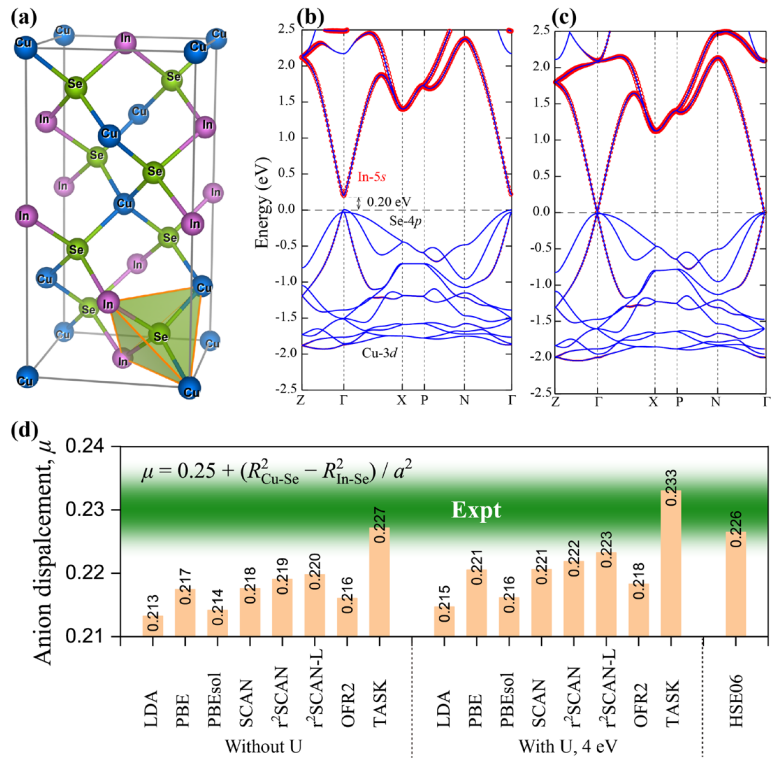

**Figure 21. Structural and electronic properties of diamond-like CuInSe$_2$.** (a) Crystal structure. The bottom tetrahedron highlights the structural unit of the diamond-like geometry. (b) r$^2$SCAN band structure using an experimental crystal structure, with $\mu = 0.230$. (c) r$^2$SCAN band structure with the r$^2$SCAN-optimized crystal structure, with $\mu = 0.219$. (d) The anion

displacement parameter, defined as $\mu = 0.25 + (R_{Cu-S}^2 - R_{Ga-S}^2)/a^2$, where $R$ and $a$ represent the bond length and lattice parameter. Experimental results exhibit considerable variation and are denoted by green-shaded areas. Subplot (d) cited from [108].

### 6.4. Reducing the remaining SIE: SCAN+$U$, FLOSIC-SCAN, and others

The SIE or delocalization error is recognized as a significant challenge for local and semi-local functionals [152], including SCAN and r$^2$SCAN. Several strategies have been developed to mitigate this error. The SCAN+$U$ method, already implemented in previous sections, serves as a straightforward yet occasionally effective solution. By penalizing the delocalization of $d$-orbitals, the $U$ term effectively reduces the SIE. As demonstrated in this paper and supported by additional studies [18,153], the $U$ values required for SCAN+$U$ are typically smaller than those for PBE+$U$. However, the +$U$ correction is not a universal solution, and the selection of $U$ values often relies on empirical methods.

Yang, Pederson, and Perdew introduced a fully self-consistent *Fermi-Löwdin orbital self-interaction correction* (FLOSIC) applicable to periodic and finite systems [154]. Subsequently, Yamamoto and colleagues implemented this approach within the SCAN functional, known as FLOSIC-SCAN, which demands much attention to numerical details and a robust integration grid to ensure reliable accuracy. FLOSIC-SCAN enhances predictions for orbital and dissociation energies, where self-interaction effects are particularly critical. However, it tends to degrade the accuracy of total and atomization energies. Despite these drawbacks, FLOSIC-SCAN generally surpasses the original SCAN functional in scenarios where SIE is significant. It remains beneficial even when SIEs are not dominant, particularly with the SCAN@FLOSIC-SCAN method [155]. This approach represents a promising route for correcting SIE efficiently in first-principles approaches, although its application to solids is limited [156].

Shahi and collaborators apply PZ-SIC to the SCAN functional [157]. This approach eliminates SIE primarily seen in stretched bonds but introduces other errors, such as orbital-density nodality. Additionally, Hui and Chai introduced a series of density functionals rooted in the SCAN framework, including SCAN0, as well as three double-hybrid functionals (SCAN0-DH, SCAN-QIDH, and SCAN0-2) without adjustable parameters [158]. Notably, SCAN0-2 consistently shows superior performance across various applications, addressing both SIE and non-covalent interaction errors without delving into static-correlation error issues, significantly enhancing the versatility of SCAN-based functionals across a broad spectrum of systems [157].

### 7. Summary

This review assesses the performance of the general-purpose SCAN and r$^2$SCAN meta-GGA functionals, explicitly focusing on transition-metal compounds. First, we introduce the design principles of the two XC functionals. Subsequently, we present a series of examples organized hierarchically, demonstrating their significant improvement over the conventional PBE GGA, primarily due to reduced self-interaction errors. Additionally, we discuss the ongoing challenges and continuous efforts to refine these functionals. This work provides comprehensive guidance to facilitate broader adoption and deeper understanding of these advanced computational tools.

### Acknowledgement

YZ is supported by the National Natural Science Foundation of China (11904156). JS acknowledge the support of the U.S. Department of Energy (DOE), Office of Science (OS), Basic Energy Sciences (BES), Grant No. DE-SC0014208.

### Conflict of Interest

The authors have no conflicts to disclose.

### Data availability

The data that support the findings of this study are available within the article.